\documentclass[aps,prx,twocolumn,superscriptaddress,showpacs,floatfix,longbibliography,UTF8]{revtex4-1}
\usepackage{amsmath,amssymb,wasysym,graphicx}

\usepackage[T1]{fontenc}
\usepackage{xcolor}

\IfFileExists{newtxtext.sty}
   {\usepackage{newtxtext,newtxmath}}
   {\IfFileExists{stix.sty}
      {\usepackage{stix}}
      {\IfFileExists{mathptmx.sty}
      {\usepackage{mathptmx}}{} } }

\usepackage{textcomp}

\usepackage{bm}
\usepackage{lipsum} 
\usepackage{float}

\IfFileExists{siunitx.sty}{\usepackage{booktabs,siunitx}}{}

\usepackage{color}
\definecolor{LinkColor}{rgb}{0.256,0.439,0.588}
\usepackage{hyperref}
\hypersetup{
   pdfauthor={guys},
   pdftitle={new title},
   colorlinks=true,
   citecolor=LinkColor,
   linkcolor=LinkColor,
   urlcolor=LinkColor
}

\renewcommand{\vec}[1]{\boldsymbol{#1}}

\usepackage{pifont}

\usepackage{multirow}
\usepackage{subfigure}

\definecolor{darkgreen}{rgb}{0.0, 0.6, 0.0}

\newcommand {\be}{\begin{eqnarray}}
\newcommand {\ee}{\end{eqnarray}}

\newcommand{\diag}{{\rm diag}}

\newcommand{\fvec}[1]{\boldsymbol{#1}}

\begin{document}
\title{Correlated Insulating Phases in the Twisted Bilayer Graphene} 

\author{Yuan Da Liao}
\affiliation{Beijing National Laboratory for Condensed Matter Physics and Institute of Physics, Chinese Academy of Sciences, Beijing 100190, China}
\affiliation{School of Physical Sciences, University of Chinese Academy of Sciences, Beijing 100190, China}

\author{Xiao Yan Xu}
\affiliation{Department of Physics, University of California at San Diego, La Jolla, California 92093, USA}

\author{Zi Yang Meng}
\email{zymeng@hku.hk}
\affiliation{Department of Physics and HKU-UCAS Joint Institute of Theoretical and Computational Physics, The University of Hong Kong, Pokfulam Road, Hong Kong SAR, China}
\affiliation{Beijing National Laboratory for Condensed Matter Physics and Institute of Physics, Chinese Academy of Sciences, Beijing 100190, China}
\affiliation{Songshan Lake Materials Laboratory, Dongguan, Guangdong 523808, China}

\author{Jian Kang}
\email{jkang@suda.edu.cn}
\affiliation{School of Physical Science and Technology \& Institute for Advanced Study, Soochow University, Suzhou, 215006, China}

\begin{abstract}
We review analytical and numerical studies of correlated insulating states in twisted bilayer graphene, focusing on real-space lattice models constructions and their unbiased quantum many-body solutions. We show that by constructing localized Wannier states for the narrow bands, the projected Coulomb interactions can be approximated by interactions of cluster charges with assisted nearest neighbor hopping terms. With the interaction part only, the Hamiltonian is $SU(4)$ symmetric considering both spin and valley degrees of freedom. In the strong coupling limit where the kinetic terms are neglected, the ground states are found to be in the $SU(4)$ manifold with degeneracy. The kinetic terms, treated as perturbation, break this large $SU(4)$ symmetry and propel the appearance of intervalley coherent state, quantum topological insulators and other symmetry-breaking insulating states. We first present the theoretical analysis of moir\'e lattice model construction and then show how to solve the model with large-scale quantum Monte Carlo simulations in an unbiased manner. We further provide potential directions such that from the real-space model construction and its quantum many-body solutions how the perplexing yet exciting experimental discoveries in the correlation physics of twisted bilayer graphene can be gradually understood. This review will be helpful for the readers to grasp the fast growing field of the model study of twisted bilayer graphene.
\end{abstract}

\date{\today}
\maketitle

\section{Introduction}
Since the discovery of correlated insulating phases and superconductivity (SC) in the twisted bilayer graphene (TBG) and other moir\'e systems near the magic angle, significant progress has been achieved in understanding the properties of the electronic correlations in these systems~\cite{bistritzer2011moire,cao2018correlated, cao2018unconventional,shen2019observation, liu2019spin, cao2019electric, chen2020tunable,kerelsky2019maximized, tomarken2019electronic, lu2019superconductors, xie2019spectroscopic, jiang2019charge, wong2019cascade, zondiner2019cascade, saito2019decoupling, stepanov2019interplay, chen2019evidence, chen2019signatures, xu2018topological, Kang2018, koshino2018,yuan2018model,Po2018PRX, liu2018chiral, ochi2018possible,dodaro2018phases,guo2018pairing,isobe2018unconventional,venderbos2018correlations,guinea2018electrostatic,liu2018pseudo,
LiuValley2019,Cea2019,tang2019spin,gonzalez2019kohn,kang2019strong,seo2019ferromagnetic,zhang2019nearly,lee2019theory,wucollective,
wu2019ferromagnetism,bultinck2019anomalous,liu2019nematic,alavirad2019ferromagnetism,chatterjee2019symmetry,chichinadze2019nematic,
bultinck2019ground,liu2019correlated,fernandes2019nematicity,zhang2020correlated,repellin2019ferromagnetism,LiuAnomalous2020,
Roy2019,Zilberberg2019,Gonzalez2017,Angeli2018,Angeli2019,2020Superconductivity,Irkhin2020,Irkhin2018Dirac,
kang2020nonabelian,huang2020slaverotor,lu2020chiral,li2019experimental,YuxuanWang2020,TianleWang2020,Christos2020,VKozii2020,WYHe2020,sharpe2019emergent,serlin2020intrinsic,TianleWang2020,VKozii2020,
xu2018kekule,YDLiao2019,fragile_topology}. The electron interaction in TBG is estimated as $e^2/(\epsilon L_m) \approx 24$meV where $\epsilon \approx 4.4$ is the dielectric constant of the hexagonal boron nitride (hBN) and $L_m \approx 13$nm is the lattice constant of the moir\'e superlattice. The bandwidth, estimated by first-principle calculations, is found to be less than $10$meV and thus smaller than the electron interactions \cite{bistritzer2011moire}, suggesting the system is either in the intermediate or strong coupling regime. While the weak coupling approach focuses on various  instabilities that are enhanced by Fermi surface (FS) nesting  and thus usually occur at incommensurate fillings, the correlated insulating states are observed only at commensurate fillings \cite{cao2018correlated,cao2018unconventional,Yankowitzeaav2019}, indicating that the electronic correlation in this system can be, at least, qualitatively understood with the strong coupling approach.

While early experiments produce the phase diagrams similar to those of heavy fermion and cuprates systems~\cite{YuanCao2019StrangeMetal}, i.e., with insulating phase, SC and strange metal above the SC dome, the discovery of the quantum anomalous Hall (QAH) state at filling number $\nu = 3$ by aligning the system with the hBN substrates \cite{sharpe2019emergent,serlin2020intrinsic} reveals the uniqueness of TBG among other strongly correlated systems. Viewing from the itinerant perspective, the band calculations produce two Dirac cones at $\fvec K$ and $\fvec K'$ of the moir\'e Brillouin zone (mBZ) for each valley~\cite{bistritzer2011moire}. Different from the graphene, the two Dirac cones have the same chirality, manifesting their non-trivial topological band properties \cite{Po2018PRX,Zou2018,song2018all,liu2018pseudo}.  As a consequence, this system is characterized by the interplay between the non-trivial topological properties and the strong interactions.  

As schematically shown in Fig.~\ref{Fig:lattice}, the two valley-polarized Dirac cones that have the same chiralities are protected by $C_2 \mathcal{T}$ symmetry where $C_2$ is the two-fold rotation around the axis perpendicular to the graphene and $\mathcal{T}$ is the time-reversal symmetry. The two nodes have the same chirality as dictated by the two valley-polarized Bloch states at the momentum of $\fvec M$ and these two Bloch states have the opposite parities under $C_{2x}$, the two fold rotation around $\hat x$ axis of Fig.~\ref{Fig:lattice} (a). Although the topological properties are not protected by adding other remote topologically trivial bands and thus dubbed as ``fragile'' topology, the two same-chirality Dirac cones cannot be reproduced by any two-band tight binding models, leading to so called topological obstruction in such non-trivial topological systems. Especially, the $C_2\mathcal{T}$ symmetry cannot be locally implemented with the Wannier states only for narrow bands~\cite{Zou2018}. Several approaches have been proposed to circumvent this difficulty, either by not locally implementing all the symmetries \cite{Po2018PRX, Kang2018, koshino2018} or including several remote bands \cite{Po2018}. In this paper, the former approach is chosen, i.e.~$C_2 \mathcal{T}$ is not locally implemented for the Wannier states, and then the associated Hamiltonian is constructed with the Coulomb interactions projected onto the narrow bands. 

Although these Dirac cones are gapped by the interaction in the correlated insulating phases, the non-trivial topological properties are crucial to have a proper understanding of the mechanism and properties of the insulating phases. Different from the conventional Hubbard model, TBG lattice models constructed based on the above consideration would acquire extended interactions and actually give rise to the emergence of ``ferromagnetic'' states due to the mechanism similar to the quantum Hall ferromagnetism at the even integer fillings, including the charge neutrality point (CNP) \cite{kang2019strong, seo2019ferromagnetic, bultinck2019anomalous, bultinck2019ground, xie2018nature, liu2019correlated, YDLiao2020}. Furthermore, the kinetic terms lift the degeneracy of the interactions and favor the inter-valley coherent state \cite{kang2019strong, bultinck2019ground,YDLiao2020}. While this order is at the momentum of $\fvec q = 0$ and thus dubbed as ``ferromagnetic'' state, it does not couple to the external magnetic field, and therefore, qualitatively consistent with the experiments.

The strongly correlated moir\'e lattice model constructed from the above principles cannot be solved analytically based on mean-field or perturbative approaches, instead, one needs to employ unbiased quantum many-body numerical calculations to achieve comprehensive understanding. Along this line, great progresses have been made from a constructive dialogue between the analytical and numerical communities. In particular, quantum Monte Carlo (QMC) simulations have been performed from carefully designed lattice models~\cite{xu2018kekule,YDLiao2019,YDLiao2020}. By implementing the fragile topology at the interaction level, one first starts with a two-band model with only spin degree of freedom but no orbital (valley) degree of freedom with cluster charge interaction on the hexagons~\cite{xu2018kekule}, the QMC simulation of this model gives rise to various translational symmetry-breaking insulating phase at CNP, for example, Kekule-type valence bond solid (VBS) phases~\cite{Lang2013}. Then with the valley degree of freedom taken into consideration, one can simulate a four-band model and investigated the interaction effect of the cluster charge repulsion upon the degenerated Dirac cones~\cite{YDLiao2019}. It is found that a continuous Gross-Neveu chiral O(4) transition happens between the Dirac cone and a VBS phase. Lastly, to fully incorporate the fragile topology in TBG, in particular the projection of the extended Wannier states onto the narrow bands, besides the cluster charge interaction, another assisted hopping interaction is introduced to the lattice model~\cite{YDLiao2020}. This turns out to be the crucial step of connecting the QMC model study with the realistic TBG experimental findings, as only with the assisted hopping on each moir\'e hexagon, can the model gives rise to intervalley coherent (IVC) and quantum valley Hall (QVH) correlated insulating phases.

What is presented in this review, is to explain in detail how the aforementioned progresses have been made in stepwise manner. We start from the theoretical considerations on what is the proper real space moir\'e lattice model of the TBG, and outline the possible correlated insulating phases suggested from the analytical calculation. Then we move on to the QMC simulation results on the lattice models inspired by the analytical considerations and follow the logic and technical flow of the numerical simulations to gradually provide the more relevant results in the ground state phase diagram at CNP. Towards the end, we will propose few immediate directions that one can take from the results presented here and make new progress, for example, the ground state phase diagram of the other integer fillings and explain what is the proper and realistic numerical tools to tackle these difficult problems.

\section{Model and Phase Diagram}
\subsection{Construction of Wannier States}
\begin{figure}[htbp]
	\centering
	\includegraphics[width=\columnwidth]{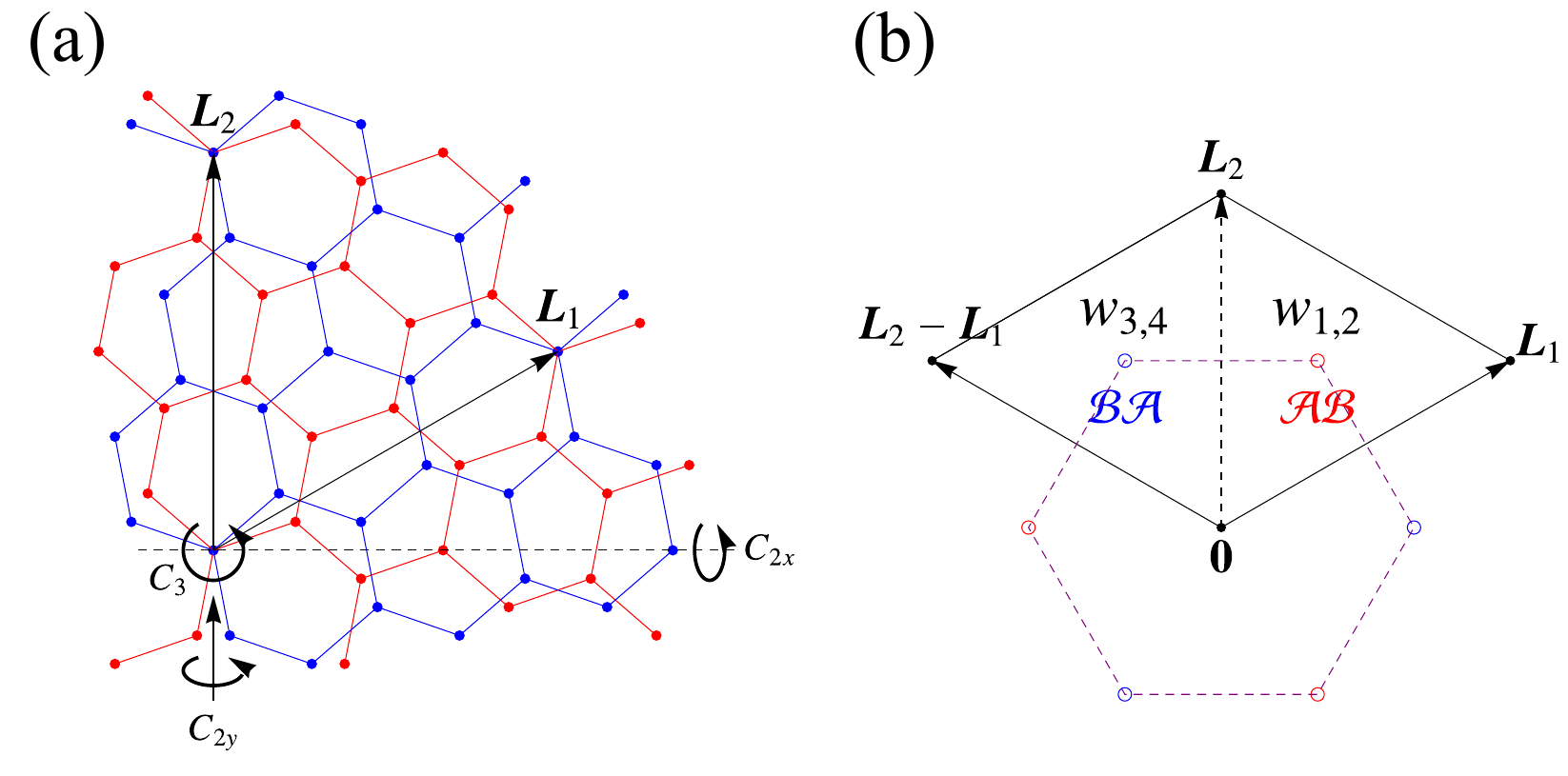}
	\caption{(a) The moir\'e superlattice structure of the TBG. Blue (red) sites are the carbon atoms on the bottom (top) layers. The triangular lattice is formed when the twisted angle is commensurate. The plot shows the lattice when the twist angle $\theta = 21.8^{\circ}$. Since the two-fold rotation $C_{2x}$ is not a symmetry transformation of the lattice, the symmetry group is $D_3$. (b) The center of the local Wannier states. Black dots are the sites of the triangular superlattice. Red and blue dots are $\mathcal{AB}$ and $\mathcal{BA}$ sites respectively, where the local Wannier states centered. In our construction, $w_1$ and $w_2$ are placed at $\mathcal{AB}$ position, and $w_3$ and $w_4$ are placed at $\mathcal{BA}$ position. Note that the $\mathcal{AB}$ and $\mathcal{BA}$ sites form an emergent honeycomb lattice. (This figure is reproduced with permission from Ref.~\cite{Kang2018}).
}
\label{Fig:lattice}
\end{figure}

In this section, we discuss the construction of the localized valley-polarized Wannier states (WS)s for the four narrow bands only. As explained above, the full symmetry of the Bitzritzer-MacDonald (BM) model~\cite{bistritzer2011moire} cannot be locally implemented because of the topological obstruction in the continuous model. To overcome this problem, we consider the discrete lattice model developed by Koshino et al.~\cite{Koshino2012}, and set the twist center axis at the registered $\mathcal{AA}$ sites. As shown in Fig.~\ref{Fig:lattice}(a), besides the time reversal symmetry, the symmetry group contains: i) the three-fold $C_3$ rotation symmetry around $\mathcal{AA}$ sites, ii) the two-fold rotation $C_{2y}$ that interchanges the layer but not the sublattice, thus forming  the $D_3$ group. Since $C_{2z}$ symmetry is not contained in $D_3$ group, this model is free of topological obstruction and all the symmetry operations of the $D_3$ group and time reversal symmetry $\mathcal{T}$ can be locally implemented.

We start from the discrete lattice model developed in Ref.~\cite{Koshino2012}
\begin{align}
H  & = -\sum_{\fvec R_i , \fvec R_j} t(\fvec r_i - \fvec r_j) f_{\fvec r_i}^{\dagger} f_{\fvec r_j} \ ,  \mbox{with} \label{Eqn:TBM_DFT} \\
t(\fvec d)  & = -V_{pp\pi} \left[ 1 - \left( \frac{\fvec d \cdot \fvec e_z}{d} \right)^2 \right] - V_{pp\sigma} \left( \frac{\fvec d \cdot \fvec e_z}{d} \right)^2 \nonumber \\
V_{pp\pi} & = V_{pp\pi}^0 \exp\left( - \frac{d - a_0}{\delta} \right) \nonumber \\
V_{pp\sigma}  & = V_{pp\sigma}^0 \exp\left( - \frac{d - a_0}{\delta} \right) \nonumber
\end{align}
where $f_{\fvec r_i}$ and $f_{\fvec r_i}^{\dagger}$ are the annihilation and creation operators of the electron at the carbon site $\fvec r_i$. Following Ref.~\cite{Koshino2012}, we set $V_{pp\pi}^0 = -2.7$eV, $V_{pp\sigma}^0 = 0.48$eV. $a_0 = 0.142$nm is the distance between the two nearest neighbor carbon atoms on the same layer.  The decay length for the hopping is $\delta = 0.319 a_0$. The hopping with $d > 4a_0$ is exponentially small and thus is neglected in the model.

Solving the tight binding model in Eq.~\eqref{Eqn:TBM_DFT}, we obtain four different narrow bands centered around the CNP. These four bands are separated from the remote bands by a band gap around $10$meV with the exact values depending on the twist angle. Although the lattice relaxation effects is important to reproduce the larger band gaps measured by the transport and STM experiments~\cite{cao2018correlated,zondiner2019cascade}, only quantitative difference is expected for the absence of this effect in our construction of WSs as our model includes only the narrow bands and the projected coulomb interaction.  Each unit cell contains four constructed WSs, labeling as $| w_{i, \fvec R_j} \rangle$ where $\fvec R_j$ labels the unit cell and $i = 1, \cdots 4$ labels the four WSs in each unit cell.

As the first step of our construction, it is crucial to identify the centers of the four WSs. One naive choice is to place them on the triangular moir\'e superlattice sites. With this option, WSs transform as
\begin{equation}
g | w_{i, \fvec R} \rangle = \sum_j | w_{j, g \fvec R} \rangle U_{ji}(g) \label{Eqn:SymmetryWannier}
\end{equation}
where $g$ is a symmetry operation in the $D_3$ group, $\fvec R_j$ specifies the position of the triangular lattice, and $g \fvec R_j$ gives the new position of the lattice site after the symmetry transformation $g$. $U(g)$ is a $4\times4$ unitary matrix that depends on $g$ and describes the transformation of the WSs. We define the Bloch state $\psi_{i, \fvec k}$ as the linear superposition of the WSs. Under the same symmetry operation $g$, we find
\begin{align}
& g | \psi_{i, \fvec k} \rangle  = g \sum_{\fvec R} e^{ i \fvec k \cdot \fvec R} |w_{i, \fvec R} \rangle = \sum_{\fvec R} e^{ i \fvec k \cdot \fvec R} |w_{j, g \fvec R} \rangle U_{ji}(g) \nonumber \\
& = \sum_{\fvec R} e^{i g\fvec k \cdot g \fvec R} |w_{j, g \fvec R} \rangle U_{ji}(g) = | \psi_{j, g \fvec k} \rangle U_{ji}(g) \ . \label{Eqn:SymWannier1}
\end{align}
It is interesting to study the special case when the momentum is symmetry invariant, i.e.~$\fvec \Gamma$ and $\fvec K$ in the mBZ. We immediately conclude that the Bloch states should transform as $U(g)$, and therefore, the Bloch states should transform in the \emph{same} way at $\fvec \Gamma$ and $\fvec K$. However, the Bloch states given by the discrete lattice model transform as two doublets at $\fvec \Gamma$ but two singlets and one doublet at $\fvec K$. This proves that the symmetry of the Bloch states cannot be reproduced if all the WSs are placed at the sites of the triangular super lattice. Further analysis shows that the symmetry can be satisfied if the centers of WSs are placed at the honeycomb lattice sites, i.e.~the centers of the WSs $| w_1 \rangle$ and $|w_2 \rangle$ at $\mathcal{AB}$ and centers of $| w_3 \rangle$ and $|w _4 \rangle$ at $\mathcal{BA}$ sites, as shown in Fig.~\ref{Fig:lattice} (b).

Once the position of the WSs are determined, the method developed by Vanderbilt~\cite{Vanderbilt2012maximally} is applied to construct the WSs, with details illustrated in Ref.~\cite{Kang2018}. Under time reversal $\mathcal{T}$, we found $| w_{1, 2} \rangle$ form a Kramer doublet, as well as $| w_{3, 4} \rangle$. Furthermore, under the symmetry transformation of the $D_3$ group
\begin{align}
C_3 | w_{1, \fvec R = 0} \rangle & = e^{i 2\pi/3}  | w_{1, - \fvec L_1} \rangle \\
C_3 | w_{2, \fvec R = 0} \rangle & = e^{-i 2\pi/3}  | w_{2, - \fvec L_1} \rangle \\
C_3 | w_{3, \fvec R = 0} \rangle & = e^{i 2\pi/3}  | w_{1, - (\fvec L_1 + \fvec L_2 )} \rangle \\
C_3 | w_{4, \fvec R = 0} \rangle & = e^{-i 2\pi/3}  | w_{2, - (\fvec L_1 + \fvec L_2)} \rangle \\
C_{2y} | w_{1, \fvec R  = 0} \rangle & = | w_{4, \fvec R = 0} \rangle \\
C_{2y} | w_{2, \fvec R  = 0} \rangle & = | w_{3, \fvec R = 0} \rangle
\end{align}
It turns out that $|w_1 \rangle$ and $| w_3 \rangle$ are contributed mostly by the states of one valley while $|w_2 \rangle$ and $| w_4 \rangle$ mostly by  the states of another valley. This leads to the different phase factor of $e^{\pm i 2\pi/3}$ under $C_3$ rotation. To be more specific, we label these WSs as $|w_{i, 1} \rangle$ and $|w_{i, 2} \rangle$, where $i$ refers to the honeycomb site, and $1$ (or $2$) specifies the valley index. For notation convenience, we introduce fermion creation/annihilation operators so that
\[ c^{\dagger}_{i, 1} | \emptyset \rangle = | w_{i, 1} \rangle \qquad  c^{\dagger}_{i, 2} | \emptyset \rangle = | w_{i, 2} \rangle \ ,  \]

\begin{figure}[htbp]
	\centering
	\includegraphics[width=\columnwidth]{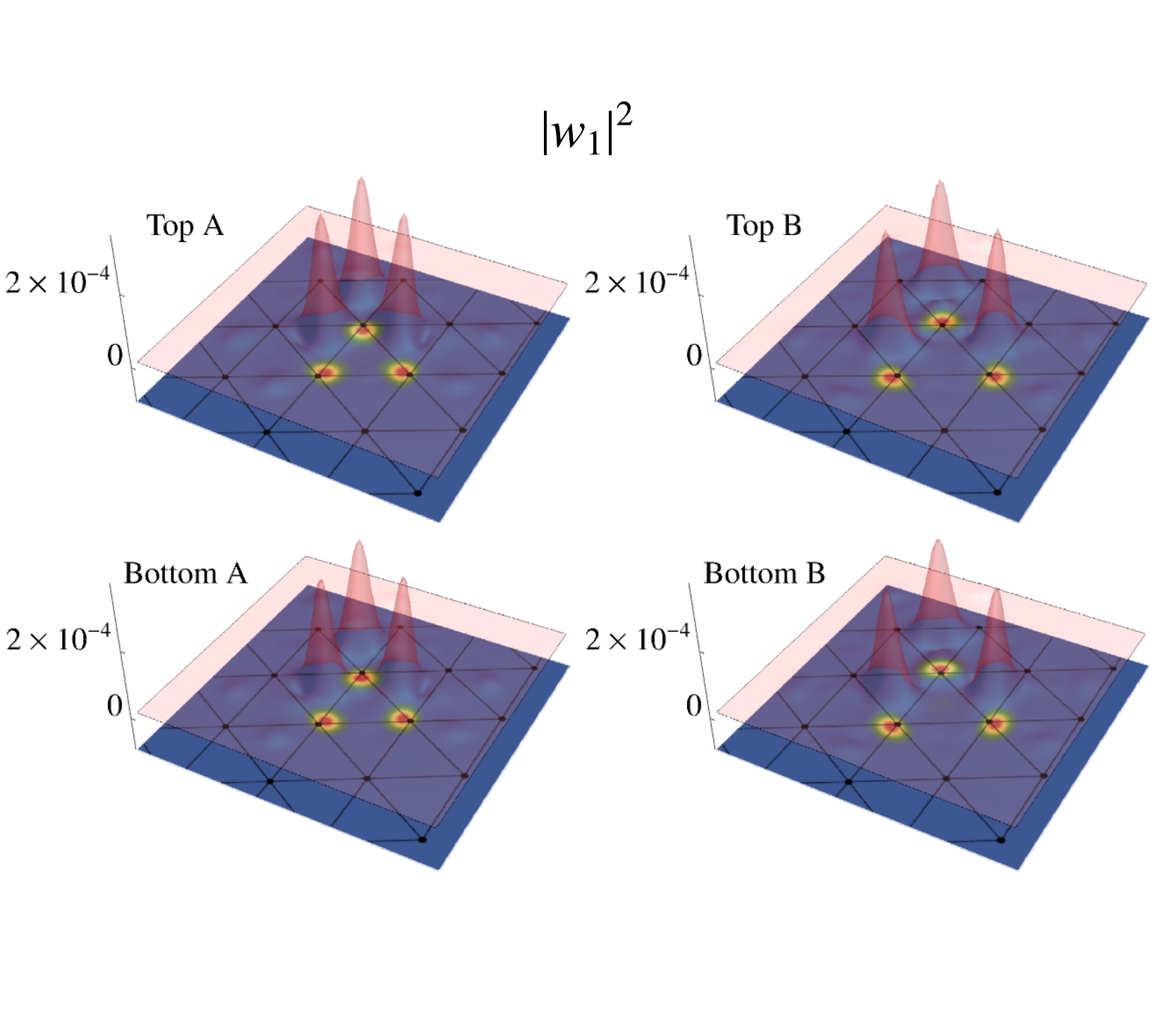}
	\caption{The localization of the WSs obtained from the projected method. The four panels show $|w_1|^2$ at (upper left) the top layer sublattice $\mathcal{A}$, (upper right) the top layer sublattice $\mathcal{B}$, (lower left) the bottom layer sublattice $\mathcal{A}$, and (lower right) the bottom layer sublattice $\mathcal{B}$. (The figure is reproduced with permission from Ref.~\cite{Kang2018}.)}
	\label{Fig:Wannier}
\end{figure}
Fig.~\ref{Fig:Wannier} shows the shape of $| w_1 \rangle$ on different layers and subattices. Each WS contains three peaks at the neighboring $\mathcal{AA}$ sites, reflected the fact that most of the LDOS are around the $\mathcal{AA}$ sites. In addition, $C_{2x}$ is not locally implemented with the WSs because it is not an exact symmetry in our discrete lattice model. More profoundly, we will see that the inability of implementing $C_{2x}$ symmetry is crucial to understand the unusual form of the interactions and how it leads to the ferromagnetic ground state. 

\subsection{The projected Coulomb interaction}
\label{sec:IIB}
Having constructed the localized WSs for narrow bands only, we project the coulomb interactions onto these WSs. As explained in Ref.~\cite{kang2019strong}, we numerically find the interaction can be well approximated by
\begin{align}
\hat H_{int} = V_0 \sum_{\varhexagon} \left( \hat Q_{\varhexagon} + \alpha \hat T_{\varhexagon} \right)^2 
\label{eq:eq10}
\end{align}
where $V_0$ is an interaction constant, depends on the dielectric constant, the gating distance, etc. The operator $\hat Q_{\varhexagon}$ counts the number of fermions located at all the six vertices of the elementary hexagon of the moir\'e superlattice, ie.
\begin{equation}
  Q_{\varhexagon} = \sum_{j = 1}^6  \left( c^{\dagger}_{j,1\sigma}   c_{j,1 \sigma} +  c^{\dagger}_{j,2\sigma}   c_{j,2 \sigma} \right)
  \label{eq:eq11}
\end{equation}
where the subscript $j = 1, \cdots, 6$ marks the hexagon sites. 

\begin{figure}[htbp]
	\centering
	\includegraphics[width=0.8\columnwidth]{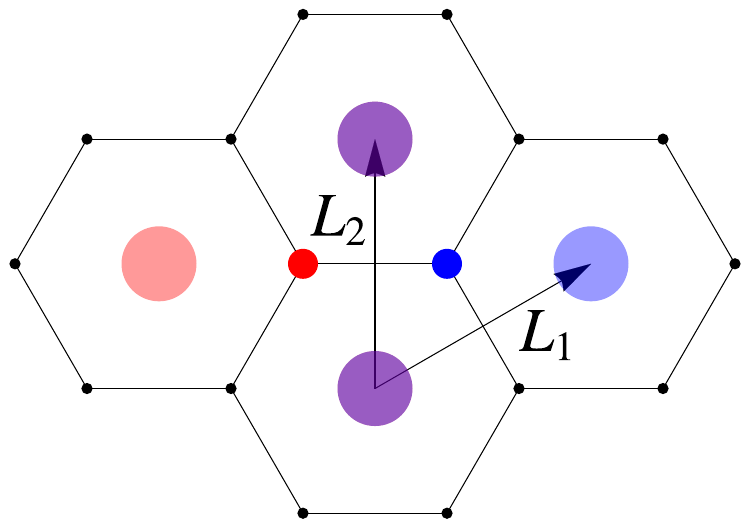}
	\caption{The centers of the hexagons correspond to the triangular moir\'e lattice spanned by primitive vectors $\boldsymbol{L}_{1,2}$. The Wannier state (WS) wavefunction centered on the moir\'e honeycomb site (black dots) has three peaks at the neighboring triangular moir\'e sites (colored disks). The overlaps of two neighboring WSs is also shown schematically with red and blue colors.}
	\label{Fig:HoneyComb}
\end{figure}

As shown in Fig.~\ref{Fig:HoneyComb}, the overlap between two neighboring WSs with the same valley contains two separated parts in two adjacent hexagons. Numerically, we found the sum of these two overlaps vanishes, as dictated by the orthogonality condition with different WSs. However, the magnitude of each overlap is $\sim O(1)$, not a small number and thus leads to the appearance of the assisted hopping term $\hat T_{\varhexagon}$ in Eq.~\eqref{eq:eq10}, given by
\begin{equation}
\hat T_{\varhexagon} \equiv \sum_{j,\sigma} \left(i c_{j+1,1\sigma}^\dagger c_{j,1\sigma} - i c_{j+1,2\sigma}^\dagger c_{j,2\sigma} + h.c. \right)
\label{eq:eq12}
\end{equation}
For notation convenience, we introduce the four-component spinor $\psi_j$ as
\[  \psi_j = \left( c_{j, 1, \uparrow},\  c_{j, 1, \downarrow},\  c_{j, 2, \uparrow},\  c_{j, 2, \downarrow}  \right)^T \ . \]
With this notation, $\hat Q_{\varhexagon} = \sum_j \psi_j^{\dagger} \psi_j$ and $\hat T_{\varhexagon} = \sum_j i \psi_{j + 1}^{\dagger} T_0 \psi_j + h.c.$, where $T_ 0 = \diag(1,\ 1,\ -1,\ -1)$. Interstingly, the interaction is invariant under a $SU(4)$ transformation:
\[  \psi_{j \in \mathcal{A}} \longrightarrow U \psi_{j} \quad \mbox{and} \quad  \psi_{j \in \mathcal{B}} \longrightarrow T_0 U T_0  \psi_{j}  \]
where $\mathcal{A}$ and $\mathcal{B}$ are two different sublattices of the honeycomb lattice. We also emphasize that only the interaction is invariant under this $SU(4)$ transformation. The whole Hamiltonian, after including the kinetic terms, breaks this $SU(4)$ symmetry, and is only valley $U(1)$ invariant.

It is worth to emphasize that this significant overlap comes from the non-trivial topological property of the narrow bands. If the narrow bands are topologically trivial,  all the symmetries would be locally implemented for the WSs, including the two-fold rotation $C_{2x}$.  Consequently, the WSs would have the same parity under $C_{2x}$, and thus, the two parts of the overlap between neighboring WSs would be the same since they are related by $C_{2x}$ symmetry. Because the sum of the two parts must vanish, each part would also vanish. This leads to the cluster Hubbard model without any assisted hopping, as shown in $H_U$ in Eqs.~\eqref{eq:model-oneorbit}, \eqref{eq:model-twoorbit} and \eqref{eq:model-twoorbitT}. The crucial role of the assisted hopping terms will be thoroughly discussed in the next section by presenting the numerical results of models with three different types of interactions.

\begin{figure*}[htp!]
	\includegraphics[width=\textwidth]{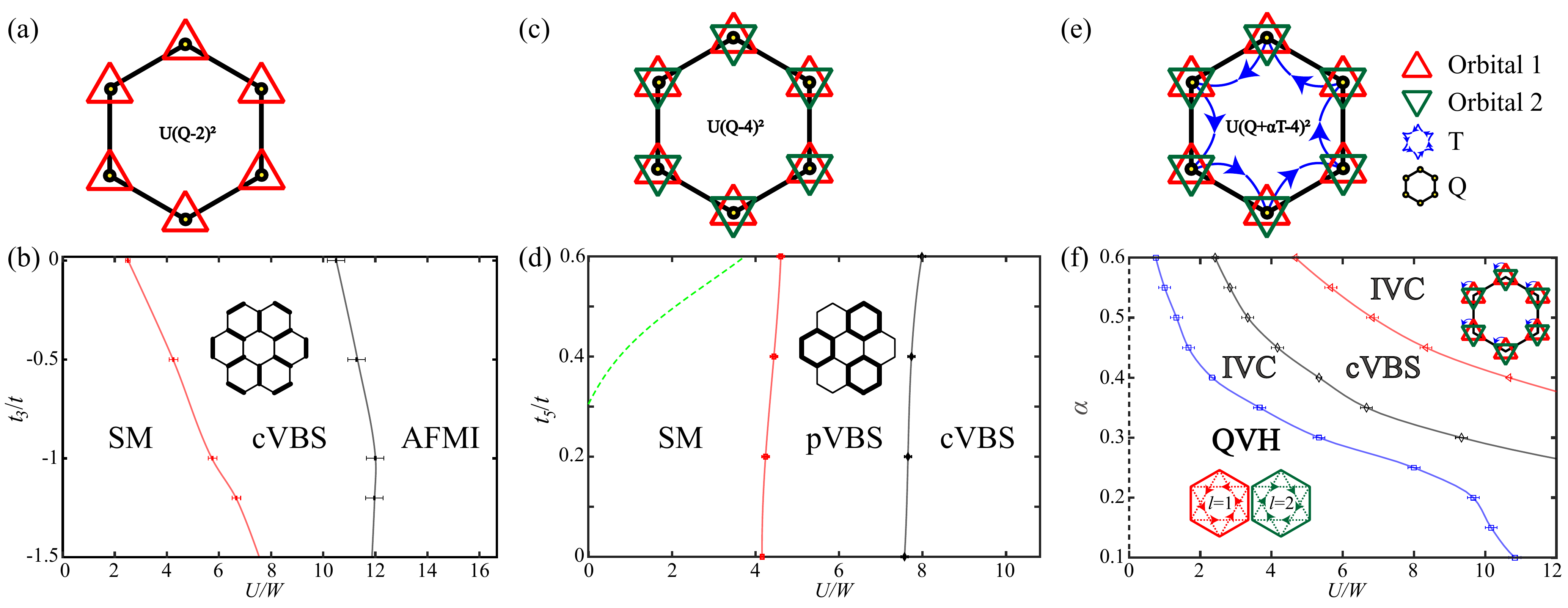}
	\caption{Honeycomb moir\'e lattice models of TBG and ground state phase diagrams at CNP obtained via QMC simulations.
		(a) Schematic representation of model described by Hamiltonian Eq.~\eqref{eq:model-oneorbit}. Here, each lattice site on the moir\'e honeycomb lattice contains one orbit (red triangles) and spins $\sigma=\uparrow,\downarrow$ (not shown). The interactions act on every hexagon and consist of the cluster charge term $Q_{\varhexagon}$ (yellow dots). 
		(b) Ground state phase diagram of (a), spanned by the $U/W$ and third-nearest neighbor hopping $-t_3/t$ axes. The transition from SM to cVBS is continuous and belongs to chiral XY universality class. The transition from cVBS to AFMI is first order. 
		(c) Schematic representation of model described by Hamiltonian Eq.~\eqref{eq:model-twoorbit}. Here, each lattice site contains two valleys $l=1,2$ (red and green triangles). 
		(d) Ground state phase diagram of (c), spanned by the $U/W$ and fifth neighbor hopping $t_5/t$ axes. The transition from SM to pVBS is continuous and belongs to chiral XY universality class. The transition from pVBS to cVBS is first order.
		(e) Schematic representation of model described by Hamiltonian Eq.~\eqref{eq:model-twoorbitT}. The interactions act on every hexagon and consist of the cluster charge term $Q_{\varhexagon}$ (yellow dots) and the assisted-hopping interaction term $T_{\varhexagon}$ (blue arrows). 
		(f) Ground state phase diagram of (e), spanned by the $U/W$ and $\alpha$ axes. The dash line at $U=0$  (y-axis) stands for the Dirac SM phase. At very small $U$, the ground state is a quantum valley Hall (QVH) phase characterized by emergent imaginary next-nearest-neighbor hopping with complex conjugation at the valley index, as illustrated by the red and green dashed hoppings with opposite directions. The system has an insulating bulk but acquires topological edge states. Upon further increasing $U$, an intervalley-coherent (IVC) insulating state is found, which breaks the SU(4) symmetry at every lattice site by removing the valley symmetry. Because it preserves the lattice translational symmetry, it is ferromagnetic-like. The columnar valence bond solid (cVBS) insulator, which appears after the IVC phase, breaks the lattice translational symmetry and preserves the onsite SU(4) symmetry. The phase transitions between QVH and IVC (blue line), between the IVC and cVBS (black line), and between the cVBS and IVC (red line) are all first order. 
		} 
	\label{fig:phasediagram}
\end{figure*}

\subsection{Honeycomb moir\'e lattice models}
Putting the above analytical considerations together and adding back the tight-bind part on a hexagonal superlattice, we can now construct the moir\'e lattice model with interaction terms in the following pedagogical steps. 

\subsubsection{One orbital (valley) model}
\label{sec:IIC1}
As discussed in our work Ref.~\cite{xu2018kekule}, following the model suggested in Po~\emph{et~al.}~\cite{Po2018PRX},   we proposed the following Hamiltonian in Eq.~\eqref{eq:model-oneorbit}, to describe the subset of hole (or electron) bands of TBG. 
\begin{equation}
\begin{aligned}
	H &= H_{0}+H_U\\
	H_{0} &=  -t\sum_{\langle ij \rangle \sigma} c^{\dagger}_{i\sigma}c_{j\sigma}+h.c. -t_3\sum_{\langle ij \rangle' \sigma}  c^{\dagger}_{i\sigma}c_{j\sigma} + h.c. \\
	H_{U} &= U\sum_{\varhexagon} (Q_{\varhexagon}-2)^2
\end{aligned}
\label{eq:model-oneorbit}
\end{equation}
where $c^{\dagger}_{i\sigma}$ ($c_{i\sigma}$) denotes creation (annihilation) operators of electrons at site $i$ with spin $\sigma=\uparrow,\downarrow$, $t=1$ is the nearest neighbor hopping on the hexagonal lattice and $t_3$ is the 3rd nearest neighbor hopping
. We use the bare bandwidth $W$ as the energy unit in the context of the paper (note that without $t_3$, the bare bandwidth $W=6t$ similar with that of the graphehe). 

As shown in Eq.~\eqref{eq:eq11}, since the WSs are quite extended in TBG, onsite, first, second and third neighbor repulsions are all important~\cite{koshino2018,Po2018PRX,kang2019strong}. To capture this kind of non-local interactions, a cluster charge Hubbard term which maintain the average filling of each elemental hexagon on the honeycomb lattice is a good choice. In Eq.~\eqref{eq:model-oneorbit}, the cluster charge $Q_{\varhexagon} \equiv \sum_{i\in \varhexagon}\frac{n_i}{3}$ with $n_i = \sum_{\sigma}c^\dagger_{i\sigma}c_{i\sigma}$ summing over all the six sites of the elemental hexagon.

This model consists of a single orbital with spin degeneracy on the honeycomb lattice. The $U/W - t_3/t$ ground state phase diagram at half-filling can be solved with QMC without sign-problem~\cite{xu2018kekule}. It is showed in Fig.~\ref{fig:phasediagram}(b). One finds three phases in total\,---\,they are semimetal (SM) phase, the AFMI phase, and a columnar valence bond solid (cVBS) phase. The transition between SM and cVBS is continuous and that from cVBS to AFMI phase appears to be first order, as will be explained in Sec.~\ref{sec:iii}.

\subsubsection{Two orbitals (valleys)  model}
\label{sec:IIC2}
In real TBG materical, there exists two valleys for the WSs, which require a four bands model with both spin and valley degrees of freedom taking into consideration. In Ref.~\cite{YDLiao2019}, we construct the following two orbital spinful lattice model on a honeycomb lattice with cluster charge interaction, 
\begin{equation}
\begin{aligned}
H &= H_{0}+H_U\\
H_{0} &=  -t\sum_{\langle ij \rangle l \sigma}c^{\dagger}_{il\sigma}c_{jl\sigma}+h.c. -t_5\sum_{\langle ij \rangle'' l\sigma}  i^{2l-1}c^{\dagger}_{il\sigma}c_{jl\sigma} + h.c. \\
H_U &= U\sum_{\varhexagon}(Q_{\varhexagon}-4)^2
\end{aligned}
\label{eq:model-twoorbit}
\end{equation}
here, orbital $l=1,2$ means the two valleys.
The fifth neighbor hopping ($it_5$ for $l=1$ and $-it_5$ for $l=2$) is purely imaginary and breaks orbital degeneracy along $\Gamma$-M direction in the high-symmetry path of BZ. 
$H_0$ is the tight-binding part introduced in Ref.~\cite{koshino2018} and serves as a minimal model to describe of the low energy band structure of TBG with Dirac points at the CNP and band splitting along $\Gamma$-M direction. The Coulomb interaction term $H_U$ is the same as the model in Eq.~\eqref{eq:model-oneorbit}, except that there are two orbits inside $Q_{\varhexagon}\equiv \sum_{i\in \varhexagon,l=1,2}\frac{n_{i,l}}{3}$.

The QMC obtianed $U/W - t_{5}/t$ ground state phase diagram at half-filling is showed in Fig.~\ref{fig:phasediagram}(d). We also found three phases in total\,---\,they are semimetal (SM) phase, a plaquette valence bond solid (pVBS) phase and a columnar valence bond solid (cVBS) phase. The two VBS are gapped insulators. Furthermore, the transition between SM and pVBS is continuous, and the phase transition from pVBS to cVBS phase appears to be first order, as will be explained in Sec.~\ref{sec:iii}.

\subsubsection{Two orbitals (valleys) model with assisted-hopping term}
\label{sec:IIC3}
As discussed in Sec.~\ref{sec:IIB}, microscopically, the full interaction of the lattice model can be derived from projecting the screened Coulomb repulsion on the narrow WS of TBG. 
Such projection leads to the emergence of an additional and sizable non-local interaction, of the form of an assisted-hopping term~\cite{kang2019strong,kang2020nonabelian}, as shown in Eqs.~\eqref{eq:eq10} and ~\eqref{eq:eq12}. As aforementioned, this new interaction ultimately arises from the non-local implementation of the $C_2 \mathcal{T}$ symmetry in a lattice model, such that the Wannier obstruction  \cite{Po2018PRX} can be overcome at the strong coupling limit. Therefore, the assisted-hopping interaction is not a simple perturbation, but a direct and unavoidable manifestation of the non-trivial topological properties of TBG.

As shown in Fig.~\ref{fig:phasediagram}(e), our model describes two valleys (orbitals) of spinful fermions on the honeycomb lattice that is dual to the triangular moir\'e superlattice. The Hamiltonian is given below, 
\begin{equation}
\begin{aligned}
H &= H_t+H_U\\
H_{0} &=  -t\sum_{\langle ij \rangle l \sigma}\left(c^{\dagger}_{il\sigma}c^{\phantom{\dagger}}_{jl\sigma}+\rm{h.c}. \right)\\
H_U &= U\sum_{\varhexagon}(Q_{\varhexagon}+\alpha T_{\varhexagon}-4)^2
\end{aligned}
\label{eq:model-twoorbitT}
\end{equation}

The two contributions of Coulomb interaction consist of the cluster charge $Q_{\varhexagon}$, which is the same as in Eq.~\eqref{eq:model-twoorbit}, and the cluster assisted hopping 
$T_{\varhexagon} \equiv \sum_{j,\sigma} \left(i c_{j+1,1\sigma}^\dagger c^{\phantom{\dagger}}_{j,1\sigma} - i c_{j+1,2\sigma}^\dagger c^{\phantom{\dagger}}_{j,2\sigma} + h.c. \right)$ as given in Eq.~\eqref{eq:eq12}. Here, the index $j=1,\ldots,6$ sums over all six sites of the elemental hexagon in the honeycomb lattice. The pre-factor $\alpha$ controls the relative strength of the two interactions. We fix the electronic filling strictly at CNP, where there are four electrons per hexagon once averaging over the lattice and the QMC can be performed without sign-problem.

The QMC-obtained phase diagram for the ground states at charge neutrality is shown in Fig.~\ref{fig:phasediagram} (f) as a function of $U/W$ and $\alpha$. 
We find that three types of correlated insulating phases emerge in the phase diagram: the quantum valley Hall (QVH) phase, the intervalley-coherent (IVC) phase, and the columnar valence bond solid (cVBS).  In the IVC phase, the two orbitals entangles with each other on every site, and thus breaks the valley $U(1)$ symmetry. The QVH appears at finite $\alpha$ and infinitesimal interaction strength $U$, and as $U$ and $\alpha$ scan through the phase diagram, the system enters into IVC, cVBS and IVC again with first order phase transition. It can also be shown that at the strong coupling limit of $U/W \to \infty$, the system is always inside an IVC phase as long as $\alpha$ is finite~\cite{YDLiao2020}. 

It is worth to emphasize that we use orbital $l=1,2$ to replace the two valleys in TBG, we believe that our second and third models in Eqs.~\eqref{eq:model-twoorbit} and ~\eqref{eq:model-twoorbitT} capture all flat bands of two valleys of TBG.

\section{Numerical Results}
\label{sec:iii}
In this section, we present the unbiased QMC results for the moir\'e lattice models in Sec.~\ref{sec:IIC1}, ~\ref{sec:IIC2} and ~\ref{sec:IIC3}.

\begin{figure}[t!]
\includegraphics[width=\columnwidth]{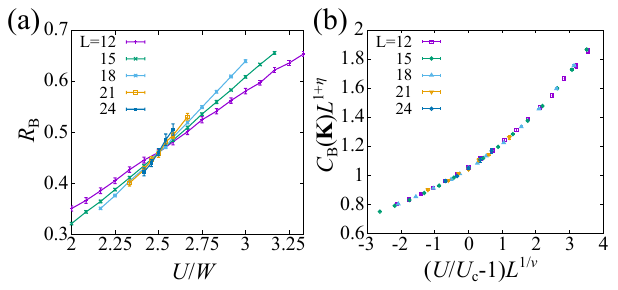}
\caption{(a) Correlation ratio of the bond correlation for $t_3/t=0$. The crossing point gives an estimate of the critical point $U_c/W=2.50(3)$.  (b) Data collapse of bond structure factor at momentum $\vec{K}$, which is the absolute value squared of the cVBS order parameter. The transition from SM to cVBS belongs to chiral XY universality class. The data collapse gives $\nu=1.05(5)$, $\eta=0.76(2)$. The figure is reproduced with permission from Ref.~\cite{xu2018kekule}.}
\label{fig:data-collapse}
\end{figure}
 
In the inset of Fig.~\ref{fig:phasediagram}(b), we present the cVBS order in real space on honeycomb lattice. It breaks the lattice translational symmetry, and the broken symmetry is $Z_3$, which will result in a signal in momentum space. To detect the signal of cVBS order, We can define a bond-bond correlation structure factor,
\begin{equation}
	C_\text{B}(\vec{q})= \frac{1} {L^4} \sum_{i,j} e^{i\vec{q}\cdot (\vec{r}_i-\vec{r}_j)} \langle B_i B_j \rangle 
\end{equation} 
where $B_i=\sum_{\alpha} ( c_{i,\alpha}^{\dagger} c_{i+\delta,\alpha} + \text{h.c.} ) $ is a bond operator. In above equations, $\delta$ means one of the three nearest-neighbour bond direction. The measurements of $C_\text{B}(\vec{q})$ show a peak at momentum points $\vec{K}$ and  $\vec{K'}$ of the first BZ.

To characterize the SM-cVBS transition, we measured the correlation ratio $R_\text{B}(U,L) =1-\frac{C_\text{B}(\vec{K}+\delta \vec{q})}{C_\text{B}(\vec{K})}$ for different system size $L$ and interaction strength $U$, where $\delta \vec{q}$ is the minimum momentum point interval of the lattice. 
This correlation ration approaches to $1$ in an ordered phase, and $0$ in the disordered one, which means the quantity is renormalization-invariant in the continuous SM-cVBS transition and will cross at a point for different system size $L$. The crossing point $U_c$ is the quantum critical point (QCP) of the SM-cVBS transition. Our numerical results are shown in Fig.~\ref{fig:data-collapse}(a), which gives an estimation of the critical point $U_c/W=2.50(3)$.

We further collapse the cVBS structure factor near the QCP with a finite size scaling relation $C_\text{B}(\vec{K},U,L)L^{z+\eta} = f_\text{B} ( (U/U_c-1)L^{1/\nu} )$, here we set the exponent $z=1$ because of Lorentz symmetry. This process can collapse all data points at one single unknown curve, as showed in Fig.~\ref{fig:data-collapse}(b). We can obtain the critical exponents $\nu=1.05(5)$ and $\eta=0.76(2)$, which are comparable with the QMC results on different models~\cite{Lang2013,zhou2016mott,scherer2016gauge,classen2017fluctuation,torres2018fermion,zerf2017four,YZLiu2020}.
 
\begin{figure}[t!]
\includegraphics[width=\columnwidth]{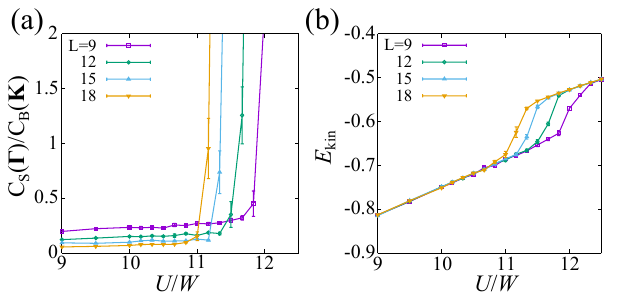}
\caption{ (a) The ratio between structure factor of AB staggered spin correlation at momentum $\boldsymbol{\Gamma}$  and  structure factor of bond correlation at momentum $\vec{K}$. The sharp jump is one evidence that cVBS to AFMI transition here is first order. (b) Kinetic energy per site of the system. Again the kink in the kinetic energy with tuning parameter $U$ is another evidence of first order transition. The figure is reproduced with permission from Ref.~\cite{xu2018kekule}.}
\label{fig:cVBS2afm}
\end{figure}

For the transition from cVBS to AFMI, we can define an AFMI structure factor $C_{\text{S}}(\boldsymbol{\Gamma})=\frac{1}{L^{4}}\sum_{ij}\langle(\vec{S}_{\mathcal{A},i}-\vec{S}_{\mathcal{B},i})(\vec{S}_{\mathcal{A},j}-\vec{S}_{\mathcal{B},j})\rangle$ to characterize it, where $\vec{S}_{\mathcal{A}/\mathcal{B},i}$ represents the spin operator of $\mathcal{A}/\mathcal{B}$ sublattice in unit cell $i$. We plot the cVBS and AFMI structure factor ration $C_{\text{S}}(\boldsymbol{\Gamma})/C_\text{B}(\vec{K})$ for different $L$. As shown in Fig.~\ref{fig:cVBS2afm}(a), this quantity  gives a singular jump, which implies that the cVBS-AFMI transition might be a first order transition. What's more, the kinetic energy per site  with tuning parameter $U$ looks like a kink, as showed in Fig.~\ref{fig:cVBS2afm}(b), which is another evidence of first order transition.

\begin{figure}[t!]
\includegraphics[width=\columnwidth]{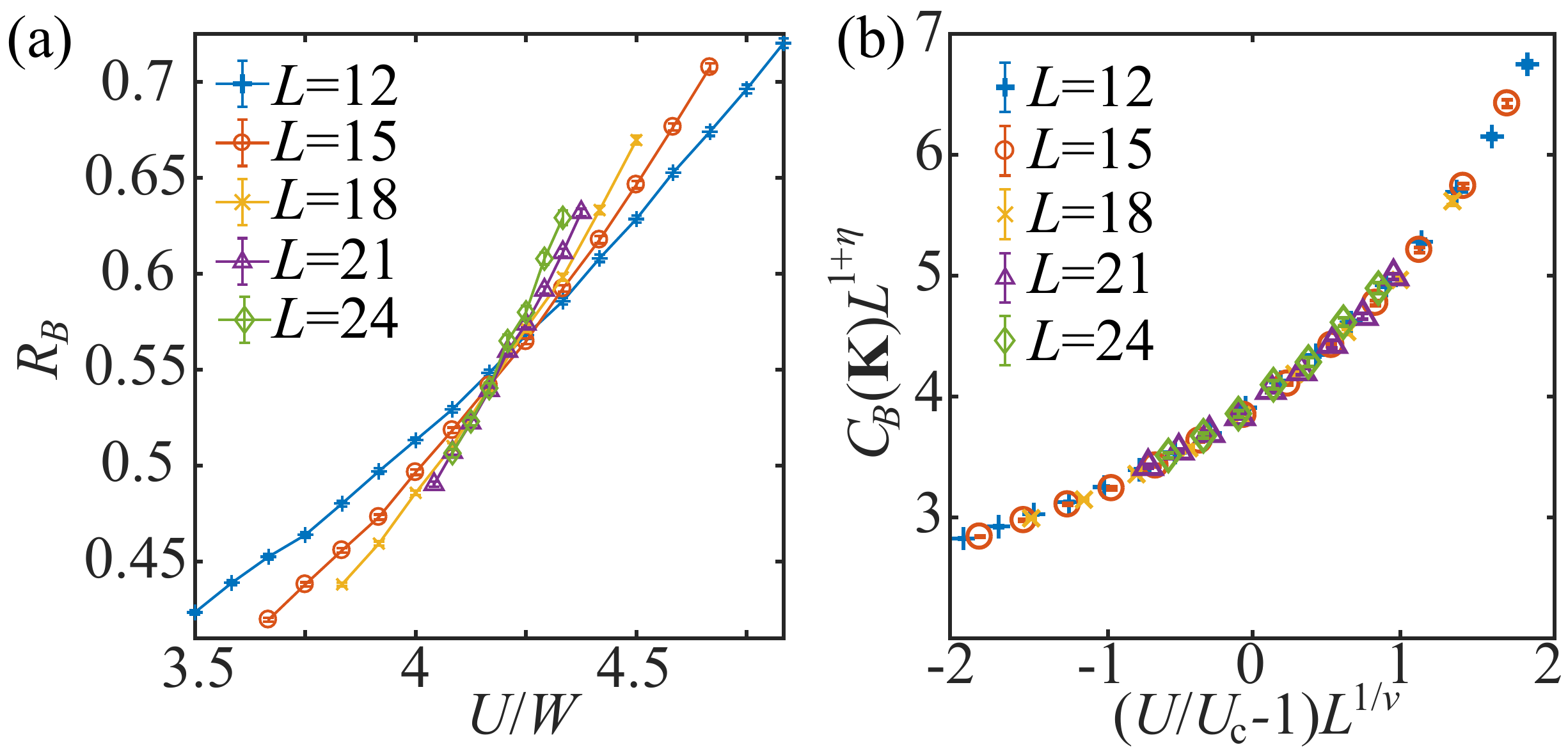}
\caption{(a) The bond-bond correlation ratio $R_{B}$ and (b) data collapse analysis of structure factor $C_{B}(\boldsymbol{K})$ at $t_5/t=0$ as function of $U/t$ with $L=12,15,\cdots,24$. The crossing of $R_{B}$ in (a) gives the DSM-pVBS critical point $U_c/t = 25.1(2)$. The data collapse in (b) gives the 3D $N=4$ Gross-Neveu chiral XY exponents $\eta=0.80(2)$, $\nu=1.01(3)$. The figure is reproduced with permission from Ref.~\cite{YDLiao2019}.}
\label{fig:crossandcollapse}
\end{figure}

The pVBS and cVBS share the same order parameter. 
For the two orbital model in Eq.~\eqref{eq:model-twoorbit}, we follow the same methodology as one orbital model to study the SM-pVBS transition. 
We also measure the bond-bond structure factor $C_B(\boldsymbol{k})$, where bond operator $B_{i,\delta}=\sum_{l,\alpha} (c_{i,l,\alpha}^\dagger c_{i+\delta,l,\alpha}+h.c.)$ with $l=1,2$. Then we plot $R_\text{B}(U,L)$ to locate the critical point of SM-pVBS transition. As shown in Fig.~\ref{fig:crossandcollapse}(a), the critical point is $U_c/W=4.18(3)$. 
We also obtain the critical exponents $\eta = 0.80(2)$ and $\nu = 1.01(3)$ from the collapse of bond-bond structure, as shown in Fig.~\ref{fig:crossandcollapse} (b). 
Because our Dirac fermions possess 4 degrees of freedom per lattice site and the pVBS phase appears an emergent $U(1)$ symmetry close to the QCP of DSM-pVBS transition as shown in Refs.~\cite{xu2018kekule,zhou2016mott}, we confirm this transition in the 3D $N=4$ Gross-Nevue chiral XY universality class~\cite{gross1974,hands1993,rosenstein1993critical,Zerf2017,li2017fermion,zhou2016mott,scherer2016gauge,mihaila2017gross,jian2017fermion,classen2017fluctuation,torres2018fermion,Ihrig2018}. 

\begin{figure}[t!]
\includegraphics[width=\columnwidth]{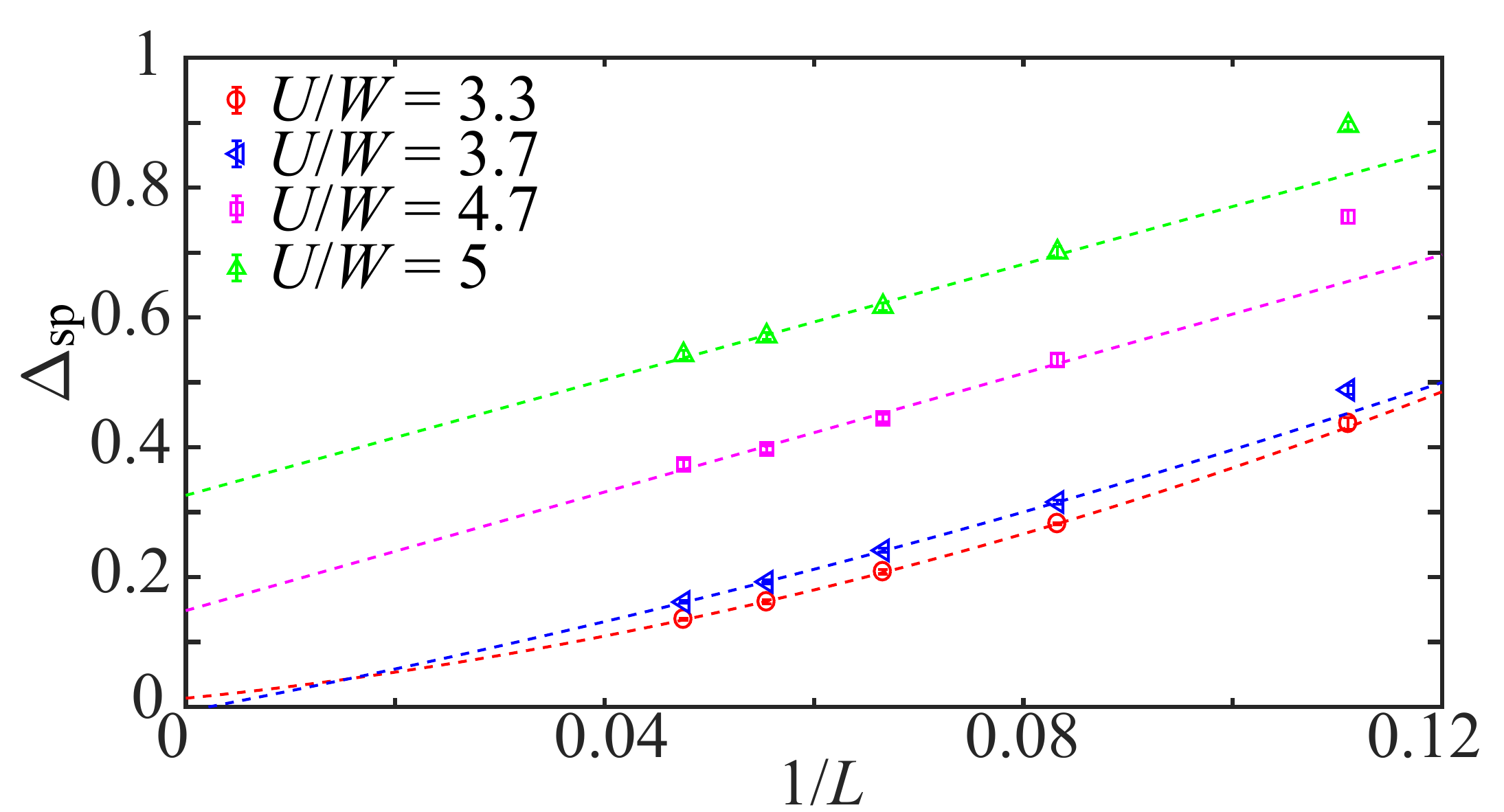}
\caption{(a) The $1/L$ extrapolation of single-particle gap $\Delta_{\text{sp}}(\boldsymbol{K})$, the gap opens between $U/t=22$ and $U/t=28$, consistent with the $U_c/t$ obtained from the bond correlation ratio in Fig.~\ref{fig:crossandcollapse} (a). The figure is reproduced with permission from Ref.~\cite{YDLiao2019}.}
\label{fig:gap}
\end{figure}

In both the phase diagrams of Fig.~\ref{fig:phasediagram} (b) and (d), the SM possesses robust massless linear dispersion at weak interaction ($U<U_c$), and the Dirac fermion will be gapped out in the insulator phase. In the one orbital model, cVBS is insulator; and in two orbital one, pVBS is also insulator. These two phase transition from SM to insulator can be monitored by measuring the dynamical single-particle Green's function. One could extract the single-particle gap from the decay relation $G(\boldsymbol{k},\tau) \propto e^{-\Delta_{\text{sp}}(\boldsymbol{k})\tau}$ at momentum $\boldsymbol{K}$, where $G(\boldsymbol{k},\tau)=\frac{1}{4L^2}\sum_{i,j,l,\sigma}e^{i\boldsymbol{k}\cdot (\boldsymbol{r}_i-\boldsymbol{r}_j)}\langle  c_{i,l,\sigma}(\frac{\tau}{2}) c^{\dagger}_{j,l,\sigma}(-\frac{\tau}{2})\rangle$. For the sake of simplicity, we only show the single-particle gap $\Delta_{\text{sp}}$ of two orbital model for different system size $L$ and interaction $U$ , as shown in Fig.~\ref{fig:gap}. It is clear that  $\Delta_{\text{sp}}\to 0$ when $U<U_c$, and $\Delta_{\text{sp}}$ goes to a finite value when $U>U_c$, suggesting the gap open at $U_c$.

%


\begin{figure}[h!]
\includegraphics[width=\columnwidth]{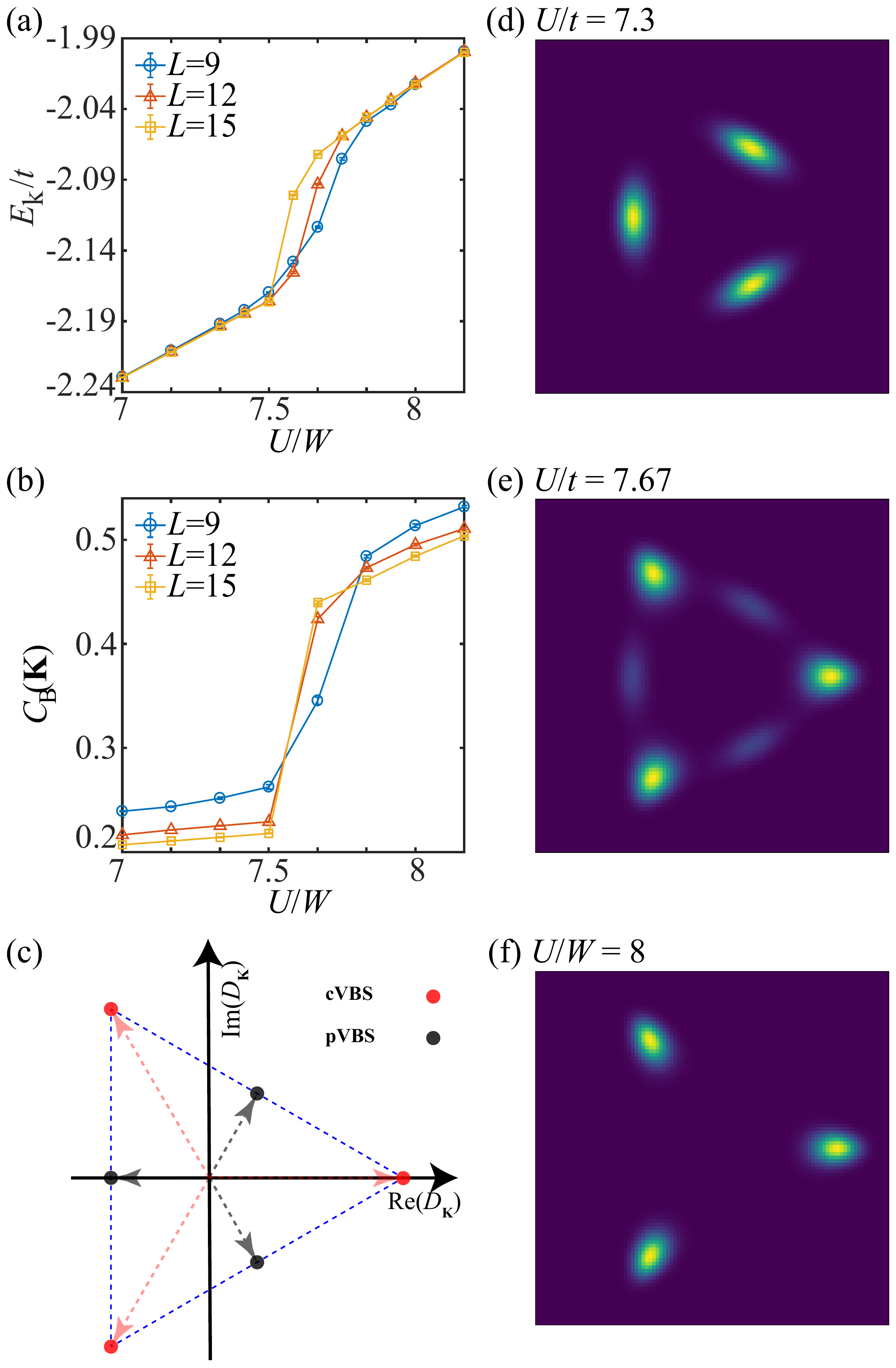}
\caption{(a) Kinetic energy per site of the system for $U$ at large values. The sharp jump signifies a first order transition. (b) $C_{B}(\boldsymbol{K})$ for the same process, a jump in VBS order is also observed, suggesting this is a transition between different VBS phases. (c) Angular dependence of the complex order parameter $D_{\boldsymbol{K}}$. Black dots represent ideal pVBS order, and red dots represent ideal cVBS order. (d)-(e) Histogram of $D_{\boldsymbol{K}}$ at different interaction strengths $U<U_{\text{VBS}}$, $U\approx U_{\text{VBS}}$ and $U>U_{\text{VBS}}$. The figure is reproduced with permission from Ref.~\cite{YDLiao2019}.}
\label{fig:secondPT}
\end{figure}

Similar with the one orbital model, if we further increase $U/W$ from the pVBS phase, we will observe a kinetic energy kink at $U/W\approx 7.67$, as shown in Fig.~\ref{fig:secondPT} (a). At the same interaction strength $U/W$ , there is also a kink of the VBS correlation, as shown in Fig.~\ref{fig:secondPT} (b). These results indicate that, a first order phase transition appears at $U_{\text{VBS}}/W \approx 7.67$ between two different VBS phases. 
There are three non-equivalent VBS configurations~\cite{zhou2016mott}, but only two of them, the pVBS and cVBS, will broke translational symmetry. In view of the fact that translational symmetry has been broken in these two VBS phase, the phase transition observed in Fig.~\ref{fig:secondPT} (a) and (b), could be the transition between pVBS and cVBS phases.

To clearly distinguish these two VBS phase, we can construct a complex order parameters $D_{\boldsymbol{K}} = \frac{1}{L^2}\sum_{i\in \mathcal{A}}\left( B_{i,\hat{e}_1}+\omega B_{i,\hat{e}_2} +\omega^2 B_{i,\hat{e}_3}\right) e^{i\boldsymbol{K}\cdot\boldsymbol{r}_i}$ following Refs.~\cite{Lang2013,zhou2016mott}, where  $\omega = e^{i\frac{2\pi}{3}}$ and $\hat{e}_1,\hat{e}_2,\hat{e}_3$ represent three nearest-neighbor bond directions. 
Theoretically, the angular distribution of  $D_{\boldsymbol{K}}$ of an ideal pVBS will point at $\frac{\pi}{3}, \pi, \frac{5\pi}{3}$, whereas that of an ideal cVBS will point at $0,\frac{2\pi}{3},\frac{4\pi}{3}$, as shown in Fig.~\ref{fig:secondPT} (c). Fig.~\ref{fig:secondPT} (d), (e) and (f) show the corresponding Monte Carlo histograms of $D_{\boldsymbol{K}}$ at three representative interaction strengths $U=7.3W < U_{\text{VBS}}$, $U=7.67W \approx U_{\text{VBS}} $, $U=8W > U_{\text{VBS}}$. 
We can notice that Fig.~\ref{fig:secondPT} (d) and (f) are inside pVBS and cVBS. Specially, Fig.~\ref{fig:secondPT}(e) clearly depicts the distribution of both characters, which is a typical example of the co-existence at the first order transition point. 

It is worth to note that the VBS phases discovered here have also been seen in $SU(N)$ Hubbard and $t-J$ models on the honeycomb lattice~\cite{Lang2013,zhou2016mott}, where only nearest neighbor hopping and on-site (Hubbard) and nearest-neighbor ($t-J$) interactions are considered. The presence of such translational symmetry breaking phase, in all these models, including the models in this work, is due to a competition between the extend interaction (or higher order processes in the multi-flavor cases even if the interaction is on-site) and the kinetic energy. When both the assisted hopping and cluster charge interactions are strong for the model in Eq.~\eqref{eq:model-twoorbitT}, the VBS phase will give way to homogeneous IVC insulators, as we will discuss below.

\begin{figure}[htp!]
\includegraphics[width=\columnwidth]{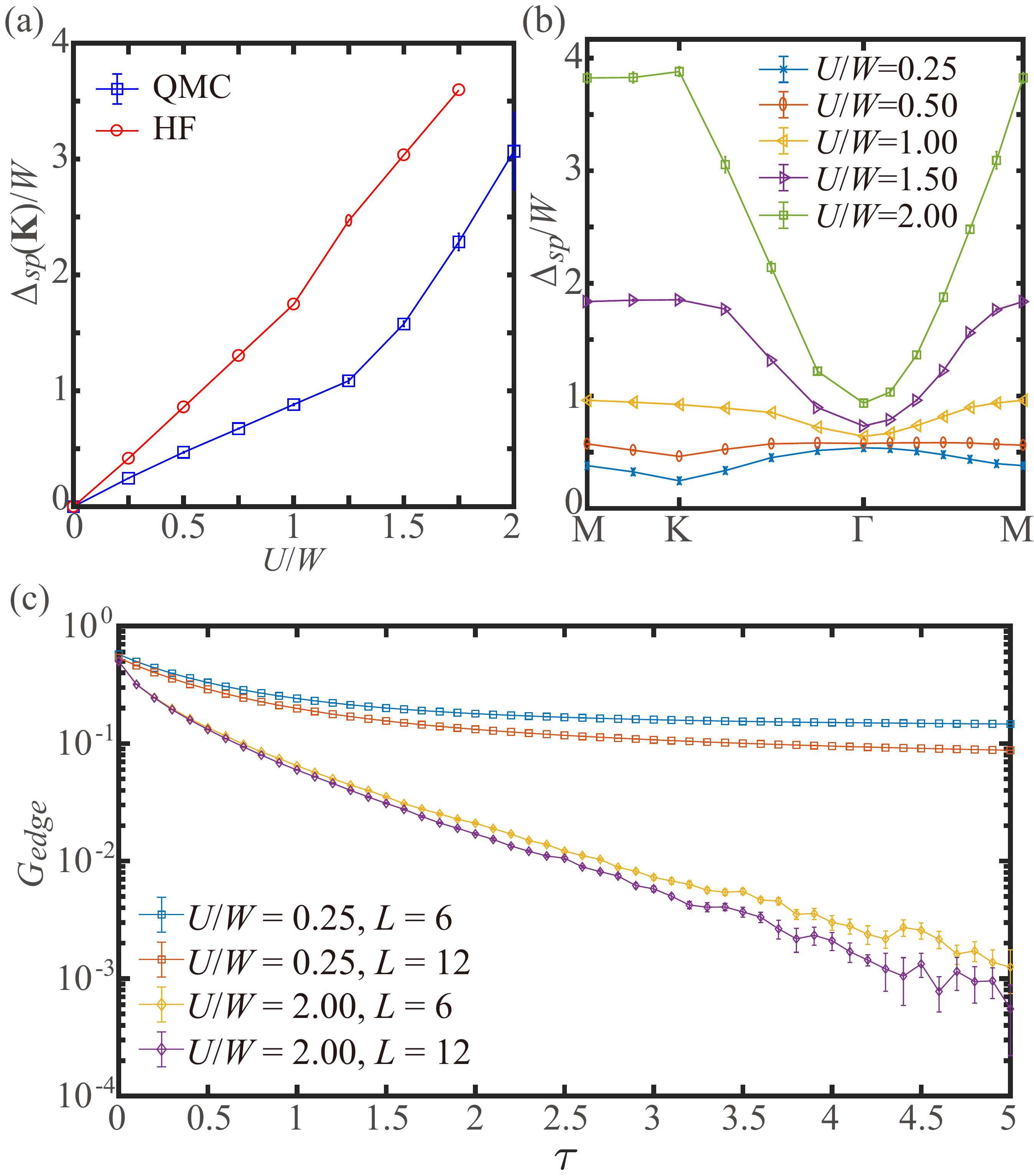}
\caption{Quantum valley Hall insulator (QVH) and gapless edge states. (a) The single-particle gap $\Delta_{\mathrm{sp}}(K)/W$ at the K point as a function of $U/W$ for $\alpha=0.45$, extracted from both QMC (blue points) and HF calculations (red points). For QMC, the spatial system size is $L=12$. The Dirac semi-metal is gapped out at the smallest small $U$ values probed. (b) Single-particle gap extracted from QMC with $L=12$ along a high-symmetry path of the Brillouin zone. (c) The topological nature of the QVH phase is manifested by valley-polarized edge states. Here we compare the edge Green's function for valley $l=1$ and spin $\uparrow$ at $U/W=0.25$ (inside the QVH phase) and $U/W=2.0$ (inside the IVC phase). It is clear that gapless edge modes only appear in the former case, highlighting the topological nature of the QVH phase. The figure is reproduced with permission from Ref.~\cite{YDLiao2020}.}
\label{fig:fig2}
\end{figure}

In many honeycomb lattice models, including the above two models, the Dirac cone at the momentum $K$ point is protected by a symmetry, and the SM is robust against relatively weak interaction strength~\cite{meng2010quantum,Lang2013,xu2018kekule,YDLiao2019,zhu2019spin}. 
Surprisingly, for the two orbital model with associated-hopping term described in Eq.~\eqref{eq:model-twoorbitT}, our QMC results revealed that a gap appeared even for the infinitesimally small values of $U$ for any $\alpha$ value that we investigated, as shown in Fig.~\ref{fig:fig2} (a). We also performed Hartree-Fock (HF) calculations on the same lattice model to verify it, the HF results, shown by the red points in Fig.~\ref{fig:fig2} (a), are in very good agreement with the QMC results. Combined with Fig.~\ref{fig:fig2} (b), it shows that the gap opens at the entire BZ at infinitesimally small $U$. 

What's more, we confirm that the gap will disappear when $\alpha=0$, in agreement with Ref.~\cite{YDLiao2019}. Together with the result that the gap appears for infinitesimally small interaction values when $\alpha \neq 0$, we could conclude that the origin of the QVH phase might come from a mean-field decoupling of the cross-term $Q_{\varhexagon} T_{\varhexagon}$. 
\begin{equation}
\sum_{\varhexagon} Q_{\varhexagon} T_{\varhexagon} = i \sum_{\varhexagon} \sum_{i, j = 1}^6 \sum_{l, m = 1}^2 (-1)^{m} \left( c^{\dagger}_{i,l} c^{\dagger}_{j+1, m} c^{\phantom{\dagger}}_{j, m} c^{\phantom{\dagger}}_{i,l} - h.c. \right)  \label{Eq:ExpandInt}
\end{equation}
where $m$ and $l$ are valley indices, spin index is omitted for simplicity. 
The terms with $j = i - 1$ and $j = i$ will cancel out after summing over all hexagons. In the weak-coupling limit, a mean-field decoupling can be performed, and there is a approximation $\langle c^{\dagger}_{i, l} c^{\phantom{\dagger}}_{i + 1, m} \rangle \propto \delta_{lm}$ because of the nearest-neighbor hopping term present in $H_0$. Then, the cross-term becomes, 
\begin{equation}
 \sum_{\varhexagon} Q_{\varhexagon} T_{\varhexagon} \propto -i \sum_{\varhexagon} \sum_{i = 1}^6 \sum_{l = 1}^2 (-1)^{l} \left( c^{\dagger}_{i, l} c^{\phantom{\dagger}}_{i + 2, l} + c^{\dagger}_{i -2, l} c^{\phantom{\dagger}}_{i, l} - h.c.  \right) \label{Eq:Cross}
 \end{equation}
Naturally, the cross-term of the interaction will induce an imaginary hopping between next-nearest-neighbors when interaction strength is relatively samll. Consequently, the mean-field Hamiltonian becomes two copies (four, if we consider the spin degeneracy) of the Haldane model \cite{Haldane88,Hohenadler2012}, resulting in a insulator with Chern number of $\pm 1$ for the two different valleys. Note that a Chern number can be defined separately for each valley $l=1$ and $l=2$ (with spin degeneracy). Because the valley $U(1)$ symmetry guarantees that these two Chern numbers must be identical, the whole system is characterized by one Chern number that takes integer values, i.e. it belongs to a $\mathcal{Z}$ classification~\cite{YYHe2016}. Due to that, we call this state as QVH phase; it is illustrated in the corresponding inset in Fig. \ref{fig:phasediagram} (f). 

It is known that there are gapless edge modes in Haldane model, despite the bulk being gapped. Form the above theoretical analysis, these edge states should be valley-polarized in the QVH phase. To verify that, we performed QMC simulations with open boundary conditions and extracted the imaginary-time Green's functions on the edge, $G_{\mathrm{edge}}(\tau)\sim e^{-\Delta_{\mathrm{sp}}\tau}$. As shown in Fig.~\ref{fig:fig2} (c), the Green's function on the edge decays to a constant in the long imaginary-time limit at small $U$ ($U/W=0.25$), the gapless edge mode disappears when increasing $U$ ($U/W=2.0$), in agreement with the theory very well
It is an important progress that the associated-hopping term qualitatively changes the ground state, as compared to the above two orbit model only with Hubbard cluster interaction. 

\begin{figure}[htp!]
\includegraphics[width=0.8\columnwidth]{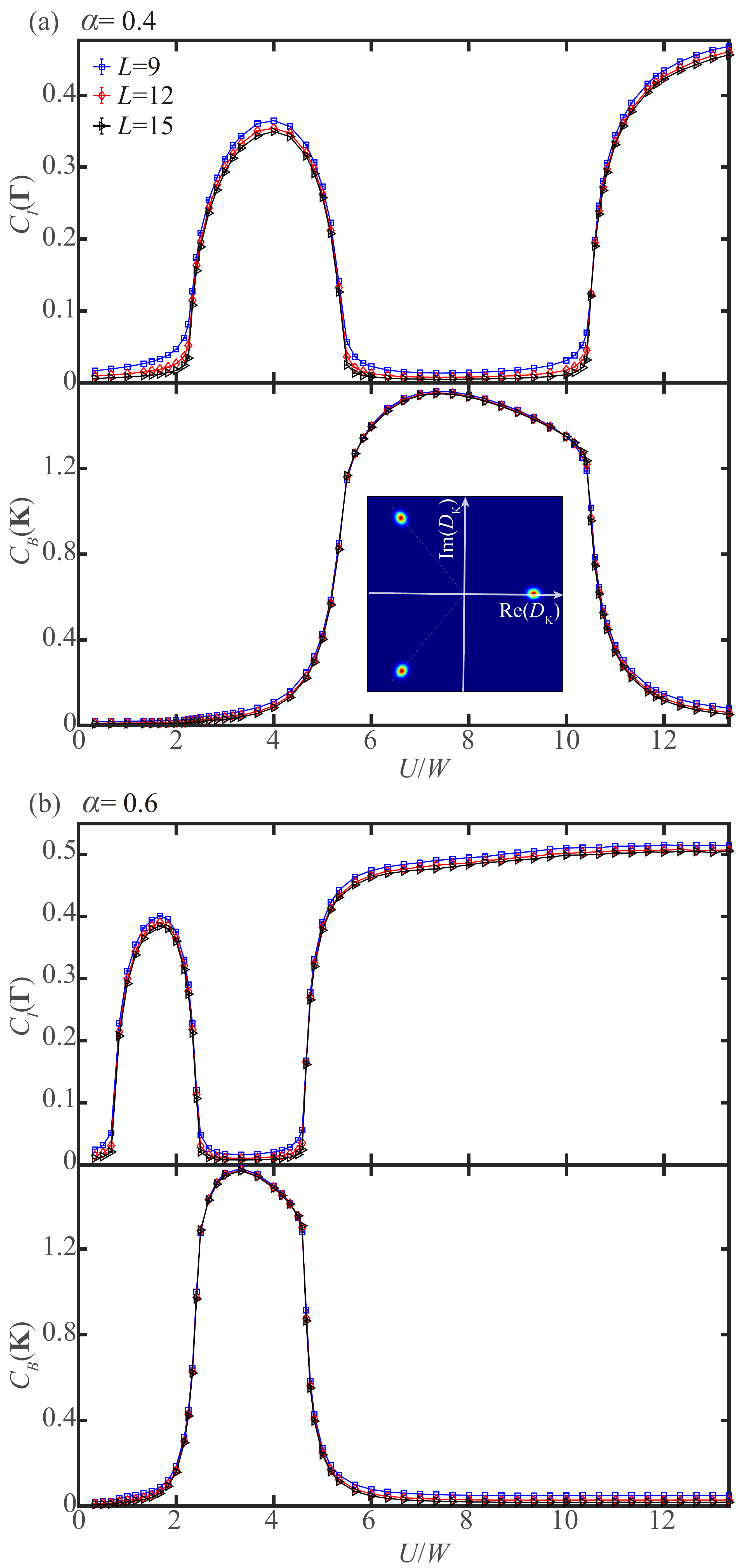}
\caption{Intervalley coherent (IVC) and columnar valence bond solid (cVBS) insulating states. Correlation functions $C_I(\boldsymbol{\Gamma})$ and $C_B(\boldsymbol{K})$, indicative of IVC and cVBS orders, respectively, as a function of $U/W$ for (a) $\alpha=0.4$ and (b) $\alpha=0.6$. Linear system sizes are indicated in the legend. In both panels, the QVH-IVC transition, the IVC-cVBS transition, and the cVBS-IVC transition are all first-order. The inset in panel (a) presents the histogram of the complex bond order parameter $D_{\boldsymbol{K}}$ at $U/W\sim5.3$. The positions of the three peaks are those expected for a cVBS phase, instead of a pVBS state. The figure is reproduced with permission from Ref.~\cite{YDLiao2020}.}
\label{fig:fig3}
\end{figure}

For larger values of $U/W$, a new insulating phase called IVC appear, which spontaneously breaks the onsite spin-valley SU(4) symmetry. To probe the IVC order, we can define a correlation function $C_I(\boldsymbol{k})=\frac{1}{L^4}\sum_{i,j \in \mathcal{A}(\mathcal{B})}e^{i\vec{k}\cdot(\boldsymbol{r}_i-\boldsymbol{r}_j)}\left\langle I_{i}I_{j} \right\rangle$, here, the operator $I_{i}=\sum_{\sigma}( c^{\dagger}_{i,l,\sigma}c^{}_{i,l',\sigma}+h.c.)$, $ l \neq l' $ represents a kind of ``onsite hopping" between two different valleys. Since there are two sublattice on honeycomb lattice, the correlation function is $C_I(\boldsymbol{k})$ a $2\times 2$ matrix, i.e. $
\begin{pmatrix}
	C^{\mathcal{A}\mathcal{A}}_I & C^{\mathcal{A}\mathcal{B}}_I \\
	C^{\mathcal{B}\mathcal{A}}_I & C^{\mathcal{B}\mathcal{B}}_I \\
\end{pmatrix}
$, these components have the relation $C^{\mathcal{A}\mathcal{A}}_I=C^{\mathcal{B}\mathcal{B}}_I=-C^{\mathcal{A}\mathcal{B}}_I=-C^{\mathcal{B}\mathcal{A}}_I$. As shown in Figs.~\ref{fig:fig3}(a) and \ref{fig:fig3}(b), we plot one of diagonal compent $C^{\mathcal{A}\mathcal{A}}_I(\boldsymbol{k})$. The correlation function is peaked at $\boldsymbol{\Gamma}$ point, which implies that the IVC order is ferromagnetic-like. The IVC order is shown in the corresponding inset of the phase diagram in Fig. \ref{fig:phasediagram} (f). Such an onsite coupling between opposite valleys breaks the valley $U(1)$ symmetry, and hence the SU(4) symmetry of the model. The fact that the SU(4) symmetry-breaking pattern is ferromagnetic-like is similar to recent analytical results \cite{kang2019strong,seo2019ferromagnetic}, which focused, however, at integer fillings away from charge neutrality.

When we further increase interaction strength $U$, as shown in Fig.~\ref{fig:fig3}, the IVC order disappear, but the cVBS structure factor at $\vec{K}$ point appear. Surprisingly, the IVC order parameter reappear when $U$ enlarge again. These numerical results are shown in Fig. \ref{fig:fig3}. 

\begin{figure}[h]
\centering
\includegraphics[width=\columnwidth]{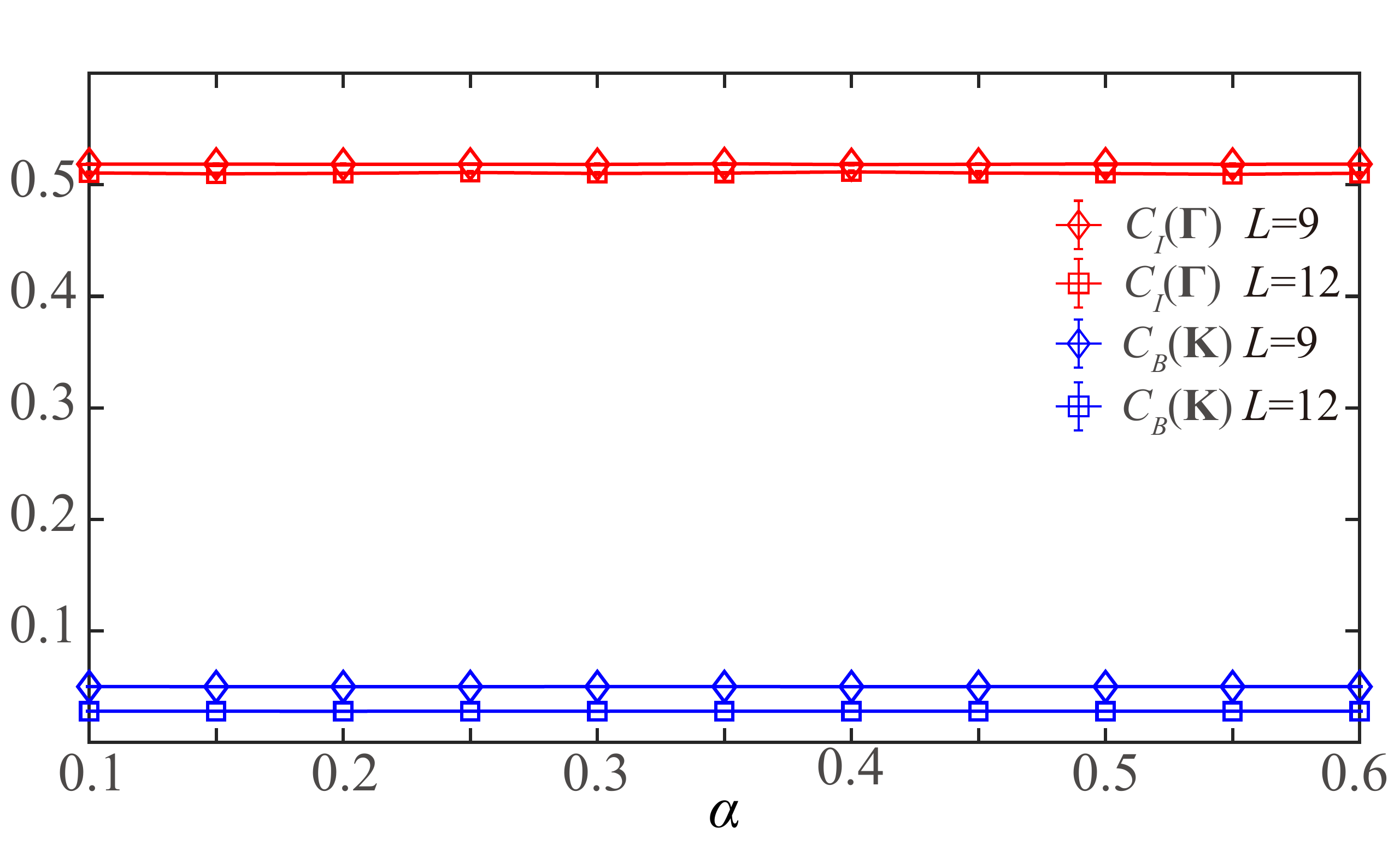}
\caption{$C_{I}(\boldsymbol{\Gamma})$ and $C_B(\boldsymbol{K})$ as a function of $\alpha$ at the strong-coupling limit. We perform the QMC simulation with projection length $\Theta=200L$, interval of time-slice $\Delta\tau=0.1$, spatial system sizes $L=9,12$, and setting $U=1$ as a dimensionless constant. The corresponding correlation function $C_{I}(\Gamma)$ of IVC is close to the saturation value of 0.5. And the correlation function $C_B(\boldsymbol{K})$ of cVBS is close to 0, which means cVBS disappear at the strong-coupling limit. The figure is reproduced with permission from Ref.~\cite{YDLiao2020}.}
\label{fig:H0}
\end{figure}

It is clear that, as $U/W$ increases, in both cases the ground state evolves from QVH to IVC to cVBS and then back to IVC. Furthermore, as will be discussed in Sec.~\ref{app:sc}, in the strong coupling limit $U/W \rightarrow \infty$, the IVC order $C_I(\boldsymbol{k} = 0)$ is independent of $\alpha$ and saturates at $0.5$. As shown in the Fig.~\ref{fig:H0}, our QMC results also confirm such expectation.

It is worth to emphasize that various translation symmetry breaking phases also appear in the numerical studies based on the continuum model. For example, recent density matrix renormalization group (DMRG) studies based on hybrid Wannier states have revealed a stripe phase, depending on the parameters of the BM model, may strongly compete with the QAH state  at the odd fillings ~\cite{kang2020nonabelian,soejima2020efficient}. In addition, the exact diagonalization (ED) based on the Bloch states also found various stripe and charge density wave (CDW) phases with the wavevector of $\mathbf{M}$ or $\mathbf{K}$ ~\cite{xie2020tbg}. Such translation symmetry breaking phases obtained by all these numerical studies, including our QMC calculation on Wannier states, establish the uniqueness of the TBG system distinguished from quantum Hall physics.

\section{Analytical Results in the Strong Coupling Limit}
\label{app:sc}
For the system at the charge neutrality point, each unit cell contains four fermions in average. Following the method applied in Ref.~\cite{kang2019strong}, the ground state $| \Psi_{gr} \rangle$ of the interaction $H_{\varhexagon}$ only should satisfy the following constraints:
\[   \hat T_{\varhexagon} | \Psi_{gr} \rangle = 0 \ ,   \]
where $T_{\varhexagon}$ is the assistant hopping terms for any hexagon. To find out the most general form of $| \Psi_{gr} \rangle$ that satisfies this constraint, we introduce the notation
\[  \psi_i' = \left( c_{i, 1, \uparrow},\ c_{i, 1, \downarrow},\ (-)^{s(i)} c_{i, 2, \uparrow},\ (-)^{s(i)} c_{i, 2, \downarrow}\right)^T \ . \]
It is obvious that the assisted hopping term $\hat T_{\varhexagon} = \sum_j i \left( \psi_j'^{\dagger} \psi'_j - h.c. \right)$. Therefore, the constraint is satisfied if and only if each honeycomb site contains exactly two ground states, and the wavefunction of the two-fermion state, in the basis of $\psi'^{\dagger}$, is identical on each site. This lead to the following form of the ground state:
\begin{align}
| \Psi_{gr} \rangle = \prod_i \sum_{\alpha \beta} U_{1 \alpha} U_{2\beta} \psi_{i, \alpha}'^{\dagger}  \psi_{i, \beta}'^{\dagger} | \emptyset \rangle \label{Eqn:FerroWaveFunc}
\end{align}
where $\psi_i' = \left( c_{i, 1, \uparrow},\ c_{i, 1, \downarrow},\ (-)^{s(i)} c_{i, 2, \uparrow},\ (-)^{s(i)} c_{i, 2, \downarrow}\right)^T$, and $U_{a \beta}$ is an arbitrary $4 \times 4$ matrix. 

We should also emphasize that not only IVC, but also valley (or spin) polarized states can be described by this general form. As a consequence of $SU(4)$ symmetry, all these states are degenerate with interactions only.

\subsection{$H_0 = 0$}
In the case of zero kinetic energy, the manifold of the ground states is described by Eqn.~\ref{Eqn:FerroWaveFunc}. To compare with the numerical results produced by QMC, we consider the correlation function:
\begin{align} 
I_{i, \uparrow} & =  \langle c^{\dagger}_{i, 1, \uparrow} c_{i, 2, \uparrow} + h.c. \rangle =  (-)^{s(i)} \left(U_{11}^* U_{13} + U_{21}^* U_{23} + c.c \right) \\
I_{i, \downarrow} & =  \langle c^{\dagger}_{i, 1, \downarrow} c_{i, 2, \downarrow} + h.c. \rangle  =  (-)^{s(i)} \left( U_{12}^* U_{14} + U_{22}^* U_{24} + c.c  \right)
\end{align}
This leads to
\begin{widetext}
	\begin{eqnarray}
	C_I^{\mathcal{A}\mathcal{A}} & =  \frac1{L^4}\sum_{i, j \in \mathcal{A}}  \langle\!\langle \left( I_{i, \uparrow} + I_{i, \downarrow} \right)  \left( I_{j, \uparrow} + I_{j, \downarrow} \right) \rangle\!\rangle  \nonumber  =  \frac1{L^4}\sum_{i, j \in \mathcal{A}}  \left( \langle\!\langle I_{i, \uparrow}  I_{j, \uparrow}  \rangle\!\rangle +   \langle\!\langle I_{i, \downarrow} I_{j, \downarrow} \rangle\!\rangle  \right)  \nonumber \\
	& =  2\times \left( |U_{11}|^2 |U_{13}|^2 + |U_{21}|^2 |U_{23}|^2 +|U_{12}|^2 |U_{14}|^2 +|U_{22}|^2 |U_{24}|^2  \right) \nonumber  = \frac12
	\end{eqnarray}
\end{widetext}
Note here $\langle\!\langle \cdots  \rangle\!\rangle$ is the average over all possible $4 \times 4$ unitary matrices and therefore $U_{ij}^* U_{k l} = \frac14 \delta_{ik} \delta_{lj}$. Similarly, we can obtain $C_I^{\mathcal{B}\mathcal{B}} = -C_I^{\mathcal{A}\mathcal{B}} = -C_I^{\mathcal{B}\mathcal{A}} = \frac12$.

\subsection{Strong Coupling Limit}
In this subsection, we assume that $H_0$ is finite but small compared with $H_{\varhexagon}$. Since the kinetic terms break $SU(4)$ symmetry, the ground state manifold shrinks and not every unitary matrix $U$ in Eqn.~\ref{Eqn:FerroWaveFunc} gives the ground state. Our purpose here is to identify the new manifold of the ground states and show that it is independent of the exact form of kinetic terms as along as they breaks the $SU(4)$ symmetry described previously.

For the convenience of calculation, we write the Eqn.~\ref{Eqn:FerroWaveFunc} as the following form, 
\begin{eqnarray}
| \psi_1 \rangle_i & = & \left( \alpha_1 c^{\dagger}_{i,1,\hat n} + (-)^{s(i)} \alpha_2 c^{\dagger}_{i, 2,\hat m}  \right) | \emptyset \rangle \\
| \psi_2 \rangle_i & = & \left( \gamma \left( \alpha_2^* c^{\dagger}_{i,1,\hat n} - (-)^{s(i)} \alpha_1^* c^{\dagger}_{i, 2,\hat m} \right) + \right. \nonumber \\
& & \left. \beta_1 c^{\dagger}_{i,1, -\hat n} + (-)^{s(i)} \beta_2 c^{\dagger}_{i,2, -\hat m}  \right) | \emptyset \rangle
\end{eqnarray}
where $s(i) = 0$ and $1$ if the site $i$ is on sublattice $\mathcal{A}$ and $\mathcal{B}$ respectively. $\hat n$ and $\hat m$ are two arbitrary spin quantization directions. $\alpha_1$, $\alpha_2$, $\beta_1$, and $\beta_2$ are four complex variables that satisfy $|\alpha_1|^2 + |\alpha_2|^2 = |\gamma|^2 + |\beta_1|^2 + |\beta_2|^2 = 1$ for normalization of the state. It is obvious that these two states are orthogonal, ie.~$\langle 1 | 2 \rangle = 0$ and also the most general form of the unentangled two-particle state on a single site $i$. 

\begin{widetext}
Therefore, The general form of the wavefunction is 
\begin{equation}  
| \Psi_{gr} \rangle  =  \prod_i \left( \alpha_1 c^{\dagger}_{i,1,\hat n} + (-)^{s(i)} \alpha_2 c^{\dagger}_{i, 2,\hat m}  \right)  \times \left( \gamma \left( \alpha_2^* c^{\dagger}_{i,1,\hat n} - (-)^{s(i)} \alpha_1^* c^{\dagger}_{i, 2,\hat m} \right) + \beta_1 c^{\dagger}_{i,1, -\hat n} + (-)^{s(i)} \beta_2 c^{\dagger}_{i,2, -\hat m}  \right) | \emptyset \rangle    
\end{equation}
Consider an arbitrary hopping between two sites. Applying the second order perturbation theory, the energy is minimized when $|\alpha_1| = |\alpha_2| = 1/\sqrt{2}$, $\gamma = 0$, and $|\beta_1| = |\beta_2| = 1/\sqrt{2}$, showing the ground state is an equal mixture of two valleys. 
\end{widetext}

To compare with the numerical result, we notice that
\begin{eqnarray} 
\langle c^{\dagger}_{i, 1, \hat n} c_{i, 2, \hat m} \rangle & = & (-)^{s(i)}  \alpha_1^* \alpha_2  \nonumber \\
\langle c^{\dagger}_{i, 1, \hat n} c_{i, 2, -\hat m} \rangle & = & \langle c^{\dagger}_{i, 1, -\hat n} c_{i, 2, \hat m} \rangle =  0 \nonumber \\
\langle c^{\dagger}_{i, 1, -\hat n} c_{i, 2, -\hat m} \rangle & = & (-)^{s(i)} \beta_1^* \beta_2  
\end{eqnarray}
Suppose that $\hat n = (\sin\theta \cos\phi, \sin\theta \sin\phi, \cos\theta)$ and $\hat m = (\sin\theta' \cos\phi', \sin\theta' \sin\phi', \cos\theta')$. We obtain that the operator
\begin{widetext}
\begin{eqnarray} 
I_{i, \uparrow} & = & \langle c^{\dagger}_{i, 1, \uparrow} c_{i, 2, \uparrow} + h.c. \rangle  =  (-)^{s(i)} \left( \cos\frac{\theta}2 \cos\frac{\theta'}2 \alpha_1^* \alpha_2 +  \sin\frac{\theta}2 \sin\frac{\theta'}2 e^{i (\phi -\phi')} \beta_1^* \beta_2  + c.c \right) \\
I_{i, \downarrow} & = & \langle c^{\dagger}_{i, 1, \downarrow} c_{i, 2, \downarrow} + h.c. \rangle = (-)^{s(i)} \left( \cos\frac{\theta}2 \cos\frac{\theta'}2 \beta_1^* \beta_2 -  \sin\frac{\theta}2 \sin\frac{\theta'}2 e^{i (\phi -\phi')} \alpha_1^* \alpha_2  + c.c  \right)
\end{eqnarray}
Since QMC simulations go through all the possible configurations of the ground states, we need to average over  $\hat n$ and $\hat m$ and thus obtain
\begin{eqnarray}
C_I^{\mathcal{A}\mathcal{A}} & = & \frac1{L^4}\sum_{i, j \in \mathcal{A}}  \langle\!\langle \left( I_{i, \uparrow} + I_{i, \downarrow} \right)  \left( I_{j, \uparrow} + I_{j, \downarrow} \right) \rangle\!\rangle  =  \frac1{L^4}\sum_{i, j \in \mathcal{A}}  \left( \langle\!\langle I_{i, \uparrow}  I_{j, \uparrow}  \rangle\!\rangle +   \langle\!\langle I_{i, \downarrow} I_{j, \downarrow} \rangle\!\rangle  \right)  \nonumber \\
& = &  2\times \left( \langle\!\langle \cos^2\frac{\theta}2 \cos^2\frac{\theta'}2 \rangle\!\rangle \left( |\alpha_1|^2 |\alpha_2|^2 + |\beta_1|^2 |\beta_2|^2 \right) + \langle\!\langle \sin^2\frac{\theta}2 \sin^2\frac{\theta'}2 \rangle\!\rangle \left( |\alpha_1|^2 |\alpha_2|^2 + |\beta_1|^2 |\beta_2|^2 \right) \right)
\end{eqnarray}
\end{widetext}
where $\langle\!\langle \cdots \rangle \!\rangle$ refers to the average over the direction $\hat n$ and $\hat m$, as well as the phases of $\alpha_1$, $\alpha_2$, $\beta_1$, and $\beta_2$. Averaging over $\hat n$ and $\hat m$ on the sphere, we obtain $\langle\!\langle \cos^2\frac{\theta}2  \rangle\!\rangle = \langle\!\langle \cos^2\frac{\theta'}2  \rangle\!\rangle =  \langle\!\langle \sin^2\frac{\theta}2  \rangle\!\rangle = \langle\!\langle \sin^2\frac{\theta'}2  \rangle\!\rangle = \frac12$. Thus, 
\[ C_I^{\mathcal{A}\mathcal{A}} = \frac12 \ .  \]
Similarly, we can obtain $C_I^{\mathcal{B}\mathcal{B}} = -C_I^{\mathcal{A}\mathcal{B}} = -C_I^{\mathcal{B}\mathcal{A}} = \frac12$.  This is consistent with the QMC result in the limit $U/W \rightarrow \infty$.

\section{Discussion}
This paper reviews the real-space lattice model construction and solution of the TBG systems and in particularly focuses on the strong coupling limit where the interactions are more important than the band structure. We first briefly explain the topological properties of the TBG material, and outline ways to circumvent the topological obstruction to construct the localized Wannier states that give rise to the narrow bands. Based on these considerations, we project the Coulomb interactions onto the narrow bands and obtain both cluster charge and assisted hopping terms as interactions in the hexagonal moir\'e lattice. We then move on to the unbiased QMC solutions of such model at CNP and summarize in a three-stage manner the phase diagrams obtained as we gradually increase the band numbers and the level of complexity in the interactions and therefore making the obtained phases more realistic and relevant with the experiments. In the strong coupling limit, the Hamiltonian at the CNP can be exactly solved and the solution is highly consistent with the QMC results. 

Our theoretical analysis and quantum many-body computation reveal the crucial role of the topological properties even in the strong coupling regime. In contrast to most strongly correlated system where the translation symmetry is usually broken, such non-trivial topological features and the strong interactions lead to the rise of the ``ferromagnetic'' orders with $\fvec q = 0$ in the TBG, or more generally, in other graphene based moir\'e systems. 
It is interesting to see from our QMC numerics, that once the assisted hopping interaction is added to the two orbital model, the IVC phase which breaks the $SU(4)$ symmetry but at the same time stay at $\fvec q=0$ is favored, amended with a QVH insulator which is symmetric and acquires non-trivial Chern number. These phases are consistent with the analytical expectations and are the very reasonable candidates for the insulating phases discovered in the TBG at even integer fillings, in particular for CNP. 

As a starting point to study the interplay between the non-trivial topological properties and the strong interactions in the electronic systems, our work provided a rather novel perspective to understand the properties of the electronic correlations. Although most of the work in this field focus on the ground states only, our numerical work also obtained the dispersion of charged excitations. It is interesting to compare this result with the STM and future ARPES experiments for more quantitative justification of our model~\cite{xie2019spectroscopic,wong2019cascade,zondiner2019cascade,liu2020spectroscopy,rozen2020entropic}.

Looking forward, here we have only revealed the QMC numerical data on the honeycomb moir\'e lattice models for TBG, and the QMC simulation is limited at CNP due to the sign problem. Another powerful method, density matrix renormalization group (DMRG), is tremendously successful in studying the low dimensional system~\cite{whitedmrg}. Recently, it was applied to study the interacting BM model without the spin and valley degrees of freedom~\cite{kang2020nonabelian,soejima2020efficient}. At the half filling where the system contain one fermion per unit cell, the numerics has identified three competing phases: QAH, strongly correlated topological semimetal, and the gapped stripe phase. While the first one breaks $C_2 \mathcal{T}$ symmetry and leads to quantized Hall conductivity, the latter two phases are $C_2\mathcal{T}$ symmetric with vanishing Hall conductivity. Although based on the simplified model and focusing only on the filling of $\nu = 1$, DMRG already produced unexpected results beyond any mean field calculations. One can expect that more sophisticated DMRG calculations and its possible combination with QMC by including spin and valley degrees of freedom and various dopings and accessing larger system sizes for the thermodynamic limit, the more complete understanding of the electronic correlations and physical mechanism behind the insulating phases, correlated metallic phase and eventually the superconducting phase in moir\'e TBG systems can be finally achieved.

\section*{Acknowledgement}
We thank Oskar Vafek, Rafael Fernandes, Clara Brei\o, Brian Andersen, Chen Shen, Guangyu Zhang, Vic Law, Xi Dai and Patrick A. Lee for the useful conversation and constructive collaborations over the projects that have been summarized in this paper. YDL and ZYM acknowledge support from the National Key Research and Development Program of China (Grant No. 2016YFA0300502) and the Research Grants Council of Hong Kong SAR China (Grant No. 17303019).  JK is supported by Priority Academic Program Development (PAPD) of Jiangsu Higher Education Institutions. YDL and
ZYM thank the Center for Quantum Simulation Sciences in the Institute of Physics, Chinese Academy of Sciences, the Computational Initiative at the Faculty of Science and the Information Technology Services at the University of Hong Kong, the Platform for Data-Driven Computational Materials Discovery at the Songshan Lake Materials Laboratory and the National Supercomputer Centers in Tianjin and Guangzhou for their technical support and generous allocation of CPU time.

\appendix

\twocolumngrid
\bibliographystyle{apsrev4-1}
\bibliography{main}
\end{document}